\documentclass[twocolumn,superscriptaddress,floatfix,preprintnumbers,prd,nofootinbib,aps,10pt]{revtex4-2}

\usepackage{graphicx,color,amsmath,amssymb}
\usepackage{slashed}
\usepackage{amsmath}
\usepackage{braket}
\usepackage[utf8]{inputenc}
\usepackage{aas_macros}
\usepackage{float}
\usepackage{bigints}
\usepackage{bbold}
\usepackage{mathtools}
\usepackage{soul}

\definecolor{navyblue}{rgb}{0.0, 0.0, 0.5}
\definecolor{royalblue}{rgb}{0.25, 0.41, 0.88}
\definecolor{cadmiumgreen}{rgb}{0.0, 0.42, 0.24}
\definecolor{blue-violet}{rgb}{0.54, 0.17, 0.89}
\definecolor{darkviolet}{rgb}{0.58, 0.0, 0.83}
\definecolor{orange(colorwheel)}{rgb}{1.0, 0.5, 0.0}

\usepackage{hyperref} 
\usepackage{amsmath}
\hypersetup{
    colorlinks=true,
    citecolor=[rgb]{.1, .7, .6},
    linkcolor=[rgb]{.2, .55, .95},
    filecolor=magenta,
    urlcolor=[rgb]{.1, .7, .6},
}

\newcommand\ie{{\it i.e.}~}
\newcommand\eg{{\it e.g.}~}

\usepackage{booktabs}
\usepackage{multirow}
\usepackage{dcolumn}
\usepackage{colortbl}

\definecolor{magenta(process)}{rgb}{1.0, 0.0, 0.56}

\definecolor{darkspringgreen}{rgb}{0.09, 0.45, 0.27}

\definecolor{royalblue(web)}{rgb}{0.25, 0.41, 0.88}

\definecolor{darkgreen}{rgb}{0.0, 0.5, 0.0}

\begin{document}

\title{Numerical Analysis of Resonant Axion-Photon Mixing: Part I}

\author{Estanis Utrilla Gin\'{e}s}
\email{estanis.ug@gmail.com}
\affiliation{Instituto de F\'{i}sica Corpuscular (IFIC), University of Valencia-CSIC, Parc Cient\'{i}fic UV, c/ Cate\-dr\'{a}tico Jos\'{e} Beltr\'{a}n 2, E-46980 Paterna, Spain}
\author{Dion Noordhuis}
\email{d.noordhuis@uva.nl}
\affiliation{GRAPPA Institute, Institute for Theoretical Physics Amsterdam and Delta
Institute for Theoretical Physics,
University of Amsterdam, Science Park 904, 1098 XH Amsterdam, The Netherlands}
\author{Christoph Weniger}
\affiliation{GRAPPA Institute, Institute for Theoretical Physics Amsterdam and Delta
Institute for Theoretical Physics,
University of Amsterdam, Science Park 904, 1098 XH Amsterdam, The Netherlands}
\author{Samuel J. Witte}
\email{samuel.witte@physics.ox.ac.uk}
\affiliation{Rudolf Peierls Centre for Theoretical Physics, University of Oxford, Parks Road, Oxford OX1 3PU, UK}

\begin{abstract}
Many present-day axion searches attempt to probe the mixing of axions and photons, which occurs in the presence of an external magnetic field. While this process is well-understood in a number of simple and idealized contexts, a strongly varying or highly inhomogeneous background can impact the efficiency and evolution of the mixing in a non-trivial manner. In an effort to develop a generalized framework for analyzing axion-photon mixing in arbitrary systems, we focus in this work on directly solving the axion-modified form of Maxwell's equations across a simulation domain with a spatially varying background. We concentrate specifically on understanding resonantly enhanced axion-photon mixing in a highly magnetized plasma, which is a key ingredient for developing precision predictions of radio signals emanating from the magnetospheres of neutron stars. After illustrating the success and accuracy of our approach for simplified limiting cases, we compare our results with a number of analytic solutions recently derived to describe mixing in these systems. We find that our numerical method demonstrates a high level of agreement with one, \textit{but only one}, of the published results. Interestingly, our method also recovers the mixing between the axion and magnetosonic-t and Alfv\'{e}n modes; these modes cannot escape from the regions of dense plasma, but could non-trivially alter the dynamics in certain environments. Future work will focus on extending our calculations to study resonant mixing in strongly variable backgrounds, mixing in generalized media (beyond the strong magnetic field limit), and the mixing of photons with other light bosonic fields, such as dark photons.
\end{abstract}

\maketitle

\section{Introduction}\label{sec:intro}
Axions are amongst the most well-motivated candidates for new fundamental physics beyond the Standard Model -- this is a result of the fact that they may play an important role in resolving fundamental problems in particle physics (namely, the so-called `strong CP problem'~\cite{PQ1,PQ2,WeinbergAxion,WilczekAxion}), they appear extensively in high energy theories~\cite{Arvanitaki:2009fg,Witten:1984dg,Cicoli:2012sz,Conlon:2006tq,Svrcek:2006yi}, and they are an ideal candidate to account for the observed abundance of dark matter in the Universe~\cite{PRESKILL1983127,Abbott1982,Fischler1982}.

One of the primary ways to search for axions is to exploit their mixing with the Standard Model photon, which arises from the dimension-five operator $\mathcal{L} \supset -\frac{1}{4} g_{a \gamma\gamma} \, a \, F_{\mu\nu} \Tilde{F}^{\mu\nu} $. Here, $a$ is the axion field, $g_{a\gamma\gamma}$ is the coupling constant (with units of inverse energy), and $F$ and $\Tilde{F}$ are the electromagnetic field strength tensor and its dual. This mixing allows for axions to convert to photons (and vice versa) in the presence of external magnetic fields, and has been employed in controlled laboratory settings (including \eg in axion haloscopes~\cite{Sikivie1983,DePanfilis:1987,Hagmann:1990,Hagmann:1998cb,Asztalos:2001tf,Asztalos:2009yp,Du:2018uak,Braine2020,Bradley2003,Bradley2004,Shokair2014,HAYSTAC,Zhong2018,Backes_2021,mcallister2017organ,QUAX:2020adt,Choi_2021,TheMADMAXWorkingGroup:2016hpc,Majorovits:2017ppy,MashaCurlyRobert2018}, axion helioscopes~\cite{CAST2009,CAST2015, Anastassopoulos2017}, and light-shining-through-walls experiments~\cite{Bibber1987,Rabadan2006,Adler2008}) as well as in astrophysical searches (see \eg~\cite{Wouters:2013hua,HESS:2013udx,Payez_2015,Fermi-LAT:2016nkz,Meyer:2016wrm,Marsh:2017yvc,Reynolds:2019uqt,Xiao:2020pra,Li:2020pcn,Dessert:2020lil,Dessert1_2022,Dessert2_2022,Noordhuis:2022ljw,Foster:2022fxn,Battye:2023oac,Janish:2023kvi,Todarello:2023hdk,Grin:2006aw,Wadekar:2021qae,Dolan:2022kul}) to set leading constraints on the coupling of axions to electromagnetism across a wide range of parameter space.

Analytic expressions characterizing the efficiency of axion-photon mixing have been derived in a variety of contexts, but are often only valid in specific limiting scenarios (such as the case of axions traversing vacuum, or axions moving in a slowly varying medium with an isotropic electromagnetic response\footnote{In this work we will frequently use the terms `isotropic' and `anisotropic' to refer to the electromagnetic response of the medium, rather than the spatial properties of the medium itself. In this context, the term isotropic thus implies that the dielectric tensor is rotationally invariant.})~\cite{Raffelt:1987im,Hook:2018iia,Battye:2019aco,Marsh:2021ajy,Battye2021,Witte2021}. This is particularly problematic for a class of searches that attempt to identify signatures arising from axion-photon mixing in extreme astrophysical environments -- here, non-trivial and/or strongly varying backgrounds can break the assumptions commonly required to derive analytic expressions, thereby dramatically complicating the mixing process\footnote{It is worth highlighting that some precision laboratory experiments, such as MADMAX, also require numerical derivations of axion-photon mixing due to the complexity of the detector geometry~\cite{knirck2019first}.}. One relevant example is a new set of ideas that uses radio telescopes to search for narrow spectral lines produced from axion dark matter resonantly converting  to photons in the magnetospheres of neutron stars~\cite{Pshirkov:2007st,Huang:2018lxq,Hook:2018iia,Safdi:2018oeu,Battye:2019aco,Leroy:2019ghm,Foster:2020pgt,Prabhu_2020,Buckley2021,Edwards:2020afl,Witte2021,Battye2021,battye2021robust,Nurmi:2021xds,Foster:2022fxn,Witte:2022cjj,Battye:2023oac,mcdonald2023generalized,Xue:2023ejt,Tjemsland:2023vvc}\footnote{A related idea is to search for excess radio emission or spectral end-points arising from axions produced locally in the magnetosphere itself~\cite{Prabhu2021,Noordhuis:2022ljw,noordhuis2023axion,Caputo:2023cpv}. These searches, however, are in general less sensitive to the uncertainties introduced by not having a complete description of axion-photon mixing.}. This is a quickly growing field which has recently demonstrated incredible potential to probe axions across a well-motivated range of parameter space, while simultaneously providing a high level of complementarity to haloscope experiments. 

The difficulty in understanding axion-photon mixing in neutron star environments boils down to a number of different effects. These include: a non-trivial electromagnetic response of the plasma (which for example may couple longitudinal and transverse modes or induce strong refractive effects, both of which can prohibit 1+1 dimensional solutions), mode-mixing, and small-scale variations in the background. While preliminary studies have attempted to incorporate the former effect (to varying degrees of success)~\cite{Hook:2018iia,Battye:2019aco,Witte2021,millar2021axionphotonUPDATED,mcdonald2023axionphoton}, the residual disagreement between these approaches have left much to be explained.

In this work, we go beyond analytic approximations and directly solve the axion-modified form of Maxwell's equations in the presence of a magnetized plasma~\cite{Wilczek:1987mv}. We compare our results with the latest analytic expressions derived in~\cite{millar2021axionphotonUPDATED,mcdonald2023axionphoton}, first confirming the validity of these expressions in weakly magnetized plasmas (where they agree), and then pushing into the regime of a strongly magnetized plasma (where the results of~\cite{millar2021axionphotonUPDATED} and~\cite{mcdonald2023axionphoton} differ strongly). We find very precise agreement with the most recent derivation of~\cite{mcdonald2023axionphoton}. This work is intended as the first in a series of papers focused on understanding axion-photon mixing in non-trivial astrophysical environments, and will provide an essential step towards accurate modeling in indirect axion searches (as an aside, we note that this approach could also prove useful for studying photon-dark photon mixing in various astrophysical settings, see \eg~\cite{McDermott:2019lch,Witte:2020rvb,Caputo:2020bdy,Caputo:2021efm,Hardy:2022ufh,Cannizzaro:2022xyw,Siemonsen:2022ivj} for a variety of applications).

The structure of this paper is as follows. We begin in Sec.~\ref{sec:method} by reviewing the axion-modified form of Maxwell's equations. We then compute the analytic axion-photon conversion probabilities that have been derived to describe the effective mixing of these particles in free space, in an isotropic plasma, and in a magnetized plasma; in the case of the latter, we also highlight the approximations under which these calculations are expected to be valid. We then introduce our numerical framework, which solves Maxwell's equations in the frequency domain. Importantly, our framework straightforwardly allows for the inclusion of non-trivial dielectric tensors, allowing us to easily transition between vacuum mixing and \eg mixing in a magnetized plasma (although the formalism is far more general, and can be applied to a wide variety of backgrounds). In Sec.~\ref{sec:results} we compare our numerical results to the analytic expressions derived in Sec.~\ref{sec:method}. Here, we show that our numerical approach can accurately reproduce free-space mixing and resonant mixing in an isotropic plasma. In the case of resonant mixing in an anisotropic plasma, we instead only find agreement with one of the published results, thereby helping to resolve an outstanding discrepancy in the field. We also comment on a number of interesting features that arise from the numerical results, including the possibility of exciting longitudinal and subluminous ($\omega < k_\gamma$) plasma modes. We conclude in Sec.~\ref{sec:conclusions}.

\section{Axion-Photon Mixing}\label{sec:method}
We begin by reviewing various analytic derivations of the axion-photon mixing probabilities under specific well-defined assumptions of the background medium. This section also outlines the numerical approach adopted in this work, which is capable of reproducing, and testing the validity of, the different analytic approximations.

\subsection{Analytic Description}
The Lagrangian describing an axion field $a$ coupled to electromagnetism is given by
\begin{multline}\label{eq:lagr}
    \mathcal{L} = -\frac{1}{4} F_{\mu\nu} F^{\mu\nu} - A_{\mu} J^{\mu} + \frac{1}{2} (\partial_\mu a \partial^\mu a - m_a^2 a^2) \\
    - \frac{1}{4} g_{a \gamma\gamma} F_{\mu\nu} \Tilde{F}^{\mu\nu} a \, ,
\end{multline}
where $A_\mu$ is the photon, $J^\mu = (\rho, \vec{J})$ is the four-current density, and $m_a$ is the mass of the axion. As mentioned in the introduction, it is the last term in Eq.~\ref{eq:lagr} that allows for axions and photons to mix. The equations of motion characterizing the evolution of the axion field, the electric field $\vec{E}$, and the magnetic field $\vec{B}$ that result from this Lagrangian are\footnote{Note that we work in units where $\epsilon_0 = c = \hbar = 1.$}
\begin{gather}
    \nabla \cdot \vec{E} = \rho - g_{a\gamma\gamma} \vec{B} \cdot \nabla a \, , \\
    \nabla \cdot \vec{B} = 0 \, , \\
    \nabla \times \vec{E} + \partial_t \vec{B} = 0 \, , \label{eq:maxfar} \\
    \nabla \times \vec{B} - \partial_t \vec{E} = \vec{J} + g_{a\gamma\gamma}(\vec{B} \partial_t a - \vec{E} \times \nabla a) \, , \label{eq:amp} \\
    (\partial_t^2 - \nabla^2 + m_a^2) a = g_{a\gamma\gamma} \vec{E} \cdot \vec{B} \, .
\end{gather}
In this work, we focus on the case of photon production from a background axion field in an astrophysical environment. Consequently, we study the above equations for the simplifying case in which a low-amplitude electromagnetic mode is excited by a background axion field passing through a cold plasma comprised only of electrons (the assumption of a cold single-species plasma is made purely for the sake of simplicity; generalizing \eg to a boosted, a thermal, or a multi-species plasma could be done via an appropriate re-definition of the dielectric tensor -- see~\cite{Witte2021} and references therein). 

Treating the plasma as a collection of individual electrons allows one to express $\rho = -e n_e$ and  $\vec{J} = -e n_e \vec{v}_e$. Here $e$, $n_e$, and $\vec{v}_e$ are the elementary charge, electron number density, and electron velocity respectively. The latter, and thereby the current density, can be obtained by solving the Lorenz force equation
\begin{equation}\label{eq:lorenz}
    m_e \partial_t \vec{v}_e = -e (\vec{E} + \vec{v}_e \times \vec{B}) \, ,
\end{equation} 
where $m_e$ is the electron mass. In order to achieve this, we assume that the electric and magnetic fields can be expanded perturbatively in powers of $g_{a\gamma\gamma}$, \ie $\vec{E} \rightarrow \vec{E}_0 + \vec{E}_1 + \dots$ and $\vec{B} \rightarrow \vec{B}_0 + \vec{B}_1 + \dots$, and that the leading order electric field in this expansion is zero, \ie $\vec{E}_0 = 0$ (as is appropriate along closed field lines in the magnetosphere, see \eg~\cite{Goldreich:1969sb,Sturrock1971a}). In a stationary background the perturbed electric and magnetic fields, as well as the velocities of electrons in the plasma (which can be expanded order by order in a manner analogous to the fields), will inherit the harmonic time dependence of the axion, which is proportional to $e^{- i \omega t}$ with frequency $\omega$. For simplicity, in the following we will drop the subscript `$1$', restricting our attention everywhere to the leading order contribution. 

Under the above assumptions, the electron velocity is given by~\cite{swanson2003plasma}
\begin{gather}
    v_{||} = \frac{-i e}{m_e \omega} \, E_{||} \, , \label{eq:vel1} \\
    v_{\perp, \pm} =  \frac{-i e}{m_e (\omega \pm \omega_c)} \, E_{\pm} \, . \label{eq:vel2}
\end{gather}
The subscripts ${||}$ and ${\perp}$ refer to the components parallel and perpendicular to $\vec{B}_0$, while the $\pm$ notation defines a rotating coordinate system in the plane perpendicular to the magnetic field, \eg $v_{\pm} = v_{\perp, 1} \pm i v_{\perp, 2}$. The cyclotron frequency is given by $\omega_c = e B_0 / m_e$. Similarly, the current density may be expressed as~\cite{swanson2003plasma}
\begin{gather}
    J_{{||}} = \frac{i \omega_p^2}{\omega} \, E_{{||}} \, , \\
    J_{\perp, \pm} = \frac{i \omega_p^2}{(\omega \pm \omega_c)} \, E_{\pm} \, ,
\end{gather}
where $\omega_p = \sqrt{e^2 n_e / m_e}$ is the plasma frequency. In what follows, we will commonly write the current density as $\vec{J} = \boldsymbol\sigma \cdot \vec{E} = i\omega \, (\mathbb{1} - \boldsymbol\epsilon) \cdot \vec{E}$. Here, $\boldsymbol{\sigma}$ is the conductivity tensor and $\boldsymbol\epsilon$ is the dielectric tensor.

We now turn our attention toward deriving a mixing equation. For the sake of completeness, we clearly outline all assumptions that enter. We begin by taking the time derivative of Ampere's law, Eq.~\ref{eq:amp}, and subsequently use the Maxwell–Faraday equation, Eq.~\ref{eq:maxfar}, to eliminate the magnetic field from the left-hand side. We then use the expansion of the electric and magnetic fields outlined above, assume harmonic time dependence for the axion and electric fields (\ie $\vec{E}(\vec{x},t) = \vec{E}(\vec{x}) e^{-i\omega t}$ and likewise for the axion), and furthermore adopt the stationary background approximation (which forces $\partial_t \vec{B}_0 = 0$). Dropping all non-leading order terms (which includes terms proportional to $g_{a\gamma\gamma} \dot{B} \dot{a}$ and $g_{a\gamma\gamma} \partial_t(\vec{E} \times \nabla a)$), one arrives at the following differential equation for the excited electric field~\cite{Hook:2018iia, Witte2021,millar2021axionphotonUPDATED,mcdonald2023axionphoton}
\begin{equation}\label{eq:waveeq}
    -\nabla^2 \vec{E} + \nabla (\nabla \cdot \vec{E}) - \omega^2 \boldsymbol\epsilon \vec{E} = \omega^2 g_{a\gamma\gamma} \vec{B}_0 a \, .
\end{equation}
We will study the above wave equation in three different backgrounds: (1) free space, which amounts to the limit $\omega_p \ll \omega$, (2) an isotropic (unmagnetized) plasma, which amounts to the limit $\omega_c \ll \omega, \omega_p$, and (3) a magnetized plasma, which amounts to the limit $\omega_c \gg \omega, \omega_p$. 

{\bf Free space -- }We first solve Eq.~\ref{eq:waveeq} for the case of an axion traversing a uniform magnetic field in vacuum, \ie $\boldsymbol\epsilon = \mathbb{1}$. Without loss of generality, we select a preferred direction for both the external magnetic field and the axion momentum -- we pick these to be the positive $x$- and $y$-directions respectively (a non-orthogonal magnetic field can be included in a straightforward manner). In free space, sourced photons do not refract, and thus all particle trajectories will stay exclusively oriented along the $y$-axis. Consequently, the electric field becomes a function only of the $y$-coordinate, and Eq.~\ref{eq:waveeq} reduces to
\begin{eqnarray}
    (\partial_y^2 + \omega^2) E_x &=& -\omega^2 g_{a\gamma\gamma} B_0 a \, , \label{eq:waveeq_free}\\
    E_y &=& 0 \, , \\
    (\partial_y^2 + \omega^2) E_z &=& 0  \, ,
\end{eqnarray}
where it should be understood that $B_0$ is, in general, spatially varying. This leaves a single equation governing the coupling of the axion to $E_x$, which can be recognized as the inhomogeneous Helmholtz equation. A general solution is obtained via the use of Green's functions, yielding 
\begin{equation}
    E_x(y) = \frac{i \omega g_{a\gamma\gamma}}{2} \int dy' e^{i \omega |y - y'|} B_0(y') a(y') \, .
\end{equation}

To continue, we define the axion-photon conversion probability $P_{a \rightarrow \gamma}$ to be the oscillation-averaged ratio of the outgoing energy flux stored in the excited electromagnetic mode(s) to the in-flowing energy stored in the axion field. The former is described by the Poynting vector $\vec{S}$, while the latter is encoded in the stress-energy tensor $T^{\mu\nu}$. We furthermore adopt a plane wave Ansatz for the axion, \ie $a(y) = a_0 e^{i k_a y}$, where $k_a$ is the axion momentum (recall that the harmonic time dependence had already been factored out in writing Eq.~\ref{eq:waveeq}). For our current case of a vacuum background, this implies
\begin{equation}\label{eq:P}
    P_{a \rightarrow \gamma} \equiv \frac{\left< S_y\right>}{\left<T^{20}\right>} = \frac{\frac{1}{2} k_\gamma \omega |A_x|^2}{\frac{1}{2} k_a \omega a_0^2}= \frac{\omega |A_x|^2}{k_a a_0^2} \, ,
\end{equation}
where we have used the free-space photon dispersion relation $\omega = k_\gamma$. Using the fact that $E_x = -i\omega A_x$ (which follows from the fact that $A_0$ depends only on the $y$-coordinate), the above leads to the final expression describing \textit{non-resonant} production of photons in vacuum 
\begin{equation}\label{eq:P_free}
    P_{a \rightarrow \gamma}^{\rm free} = \frac{\omega g_{a\gamma\gamma}^2}{4 k_a} \left| \int dy' e^{i \omega |y - y'|} e^{i k_a y'} B_0(y') \right|^2 \, .
\end{equation}
In the limit $y \rightarrow \infty$ (\ie in the limit where the axion/photon has left the magnetic field region), this integral is simply the Fourier transform of the magnetic field $\tilde{B}(k)$ evaluated at $k = k_\gamma - k_a$. This is a well-known result~\cite{Raffelt:1987im,Sigl:2017sew}, which reflects the fact that the production of an on-shell photon from an on-shell axion requires a non-zero energy/momentum contribution from the magnetic field, which naturally arises from the breaking of the translational invariance imposed by the boundary of the magnetic field region.

{\bf Isotropic plasma -- }We next turn our attention to the case of \textit{resonant} photon production from an isotropic, unmagnetized plasma (recall that the term `isotropic' is used to describe the local response of the plasma, not the medium itself). This limit is obtained by taking $\omega_c \ll \omega, \omega_p$, and yields a dielectric tensor of the form $\boldsymbol\epsilon = \textrm{diag}(1 - \omega_p^2 / \omega^2, 1 - \omega_p^2 / \omega^2, 1 - \omega_p^2 / \omega^2)$~\cite{swanson2003plasma}, which is manifestly rotationally invariant. The electromagnetic modes can be obtained by solving for the Eigenmodes of the wave equation in the absence of an axion source term. Approximating $\partial_i E_j = k_i E_j$, this results in the corresponding dispersion relations being given by the roots of $|| n^2 \delta_{ij} - n_i n_j - \epsilon_{ij}||$ (where $|| \cdot ||$ means the determinant), where $n_i = k_i / \omega$ is the refractive index. For an isotropic plasma, this procedure yields a single non-trivial dispersion relation $\omega^2 = k_\gamma^2 + \omega_p^2$\footnote{One of the roots of this equation does yield an additional longitudinal mode with dispersion relation $\omega = \omega_p$.} -- notice that this looks much like the dispersion relation of a free massive particle, where the plasma frequency plays the role of an effective mass.

The presence of an effective mass contribution opens up the possibility for the photon and axion four-momentum to be exactly equal, \ie $k^\mu_\gamma = k^\mu_a$. In an isotropic plasma this occurs when the plasma frequency is equal to the axion mass, \ie when $\omega_p = m_a$. At this `resonance', energy-momentum conservation is naturally satisfied. Unlike the vacuum mixing scenario described above, the resonant contribution to photon production does not depend on the spatial inhomogeneity of the magnetic field region, and in many cases provides the dominant contribution to photon production -- this is for example true for axion-photon mixing near neutron stars. Accordingly, in this paper we will always work in the limit where the resonance, if permitted, dominates photon production.

Plugging the isotropic dielectric tensor into Eq.~\ref{eq:waveeq} yields
\begin{gather}
    \multlinegap=71pt
    \begin{multlined}
        \partial_y^2 E_x - \partial_x\partial_y E_y + (\omega^2 - \omega_p^2) E_x \\
        = - \omega^2 g_{a\gamma\gamma} B_0 \sin\theta_B a \, ,
    \end{multlined} \\
    \multlinegap=71pt
    \begin{multlined}
        \partial_x^2 E_y - \partial_y\partial_x E_x + (\omega^2 - \omega_p^2) E_y \\
        = - \omega^2 g_{a\gamma\gamma} B_0 \cos\theta_B a \, ,
    \end{multlined}
\end{gather}
where we have again confined ourselves to the $x$-$y$ plane, resulting in the decoupling of $E_z$, which can be done while maintaining generality. The axion momentum has furthermore been chosen to point in the positive $y$-direction, and the external magnetic field to lie in the first quadrant, rotated at an angle $\theta_B$ from the $y$-axis (see Fig.~\ref{fig:simdetails}). Assuming the magnetic field varies only in the $y$-direction, the WKB limit (in which the wavelength of the axion and photon are short with respect to the scale over which the background varies) ensures that the electric field does not depend on the $x$-coordinate near the resonance, and thus the mixing equations there reduce to
\begin{eqnarray}\label{eq:waveeq_isoplasma}
    (\partial_y^2 + \omega^2 - \omega_p^2) E_x &=& -\omega^2 g_{a\gamma\gamma} B_0(y) \sin \theta_B a \, , \\
    E_y &=& \frac{-\omega^2 g_{a\gamma\gamma} B_0(y) \cos\theta_B a}{\omega^2 - \omega_p^2} \, ,
\end{eqnarray}
where we have explicitly expressed the $y$-dependence of the magnetic field. Notice that the first equation describes on-shell photon production while the second reflects the axion-induced electric field, our interest here being in the former.

While Eq.~\ref{eq:waveeq_isoplasma} looks very similar to Eq.~\ref{eq:waveeq_free}, the plasma frequency has a positional dependence, for which there is no analytic expression for the Green's function. Instead we adopt the Ansatz
\begin{equation}\label{eq:E_iso_ansatz}
    E_x(y) = \tilde{E}_x(y) e^{i k_a y} \, ,
\end{equation}
which, under the WKB approximation, allows one to write Eq.~\ref{eq:waveeq_isoplasma} as
\begin{equation}
    2 i k_a \partial_y \tilde{E}_x \simeq -(m_a^2 - \omega_p^2) \tilde{E}_x -\omega^2 g_{a\gamma\gamma} B_0 \sin \theta_B a_0 \, .
\end{equation}
This can be recognized as a Schr\"{o}dinger-like equation, which has the solution
\begin{equation}
    \tilde{E}_x(y) \simeq \frac{i \omega^2 g_{a\gamma\gamma} a_0}{2 k_a} \int_{-\infty}^y dy' B_0(y') \sin\theta_B(y') e^{i \phi(y')} \, ,
\end{equation}
where the phase is given by
\begin{equation}
    \phi(y) = \int_0^y dy' \frac{m_a^2 - \omega_p^2(y')}{2 k_a} \, .
\end{equation}
Using the stationary phase approximation (which relies on the assumption that the scale of the mixing is large when compared to the width of the stationary phase), the electric field amplitude asymptotically far from the resonance is given by
\begin{equation}
    \tilde{E}_x \simeq \frac{i \omega^2 g_{a\gamma\gamma} a_0}{2 k_a} B_0(y_c) \sin \theta_B(y_c) e^{i \phi(y_c) + \frac{i \pi}{4}} \sqrt{\frac{2 \pi}{|\partial_y^2 \phi (y_c)|}} \, .
\end{equation}
Here, $y_c$ denotes the $y$-coordinate at the resonant conversion point.

Defining the conversion probability analogously to Eq.~\ref{eq:P}, and evaluating the Poynting flux asymptotically far from the resonance, one arrives at
\begin{equation}\label{eq:prob_iso}
    P_{a \rightarrow \gamma}^\textrm{iso} = \frac{\pi}{2} \frac{\omega^2 g_{a\gamma\gamma}^2 B_0^2 \sin^2 \theta_B}{\omega_p k_a} \frac{1}{|\partial_y \omega_p|} \, ,
\end{equation}
where all relevant quantities are understood to be evaluated at the resonance -- this is again, a well-known result~\cite{Raffelt:1987im} (see also~\cite{Battye:2019aco} for a generalized derivation to arbitrary geometries). The conversion probability thus only depends on the amplitude of the magnetic field at the resonance (rather than the Fourier transform), and moreover contains a contribution $\propto |\partial_y \omega_p(y_c)|^{-1}$ describing how quickly $\omega_p$ departs from $m_a$ as the photon moves away from the conversion point (\ie it captures the effective de-tuning of the axion and photon dispersion relations away from the resonance).

{\bf Anisotropic plasma -- }In the limit of a very strong magnetic field, $\omega_c \gg \omega, \omega_p$, Eqs.~\ref{eq:vel1} and~\ref{eq:vel2} show that electrons are confined to flow exclusively in the direction of this intense magnetic field. As a result, an electromagnetic mode propagating through a magnetized medium will only excite plasma oscillations along the axis of the magnetic field; that is to say, the plasma response is anisotropic, having a preferred direction in space. Assuming a magnetic field confined to the $x$-$y$ plane (rotated from the $+y$-axis toward the $+x$-axis by an angle $\theta_B$), the dielectric tensor can be expressed as~\cite{swanson2003plasma} 
\begin{equation}\label{eq:dielectrictensor}
    \boldsymbol\epsilon = \left(\begin{array}{cccc}
    1 - \frac{w_p^2}{w^2} \sin^2 \theta_B & -\frac{w_p^2}{w^2} \sin \theta_B \cos \theta_B & 0 \\
    -\frac{w_p^2}{w^2} \sin \theta_B \cos \theta_B & 1 - \frac{w_p^2}{w^2} \cos^2 \theta_B  & 0 \\
    0 & 0 & 1
    \end{array}\right) \, ,
\end{equation}
This definition of the dielectric tensor yields three distinct electromagnetic modes, one of which is superluminal ($\omega > k_\gamma$) and two of which are subluminal ($\omega < k_\gamma$)~\cite{Witte2021, millar2021axionphotonUPDATED, mcdonald2023axionphoton}. These include the Langmuir-O (LO) mode
\begin{equation}
    \omega^2 = \frac{1}{2}\left(k_\gamma^2 + \omega_p^2 + \sqrt{k_\gamma^4 + \omega_p^4 - 2k_\gamma^2 \omega_p^2 \cos(2 \theta_{kB})}\right) \, , 
\end{equation}
the Alfv\'{e}n mode
\begin{equation}
    \omega^2 = \frac{1}{2}\left(k_\gamma^2 + \omega_p^2 - \sqrt{k_\gamma^4 + \omega_p^4 - 2k_\gamma^2 \omega_p^2 \cos(2 \theta_{kB})}\right) \, ,
\end{equation}
and the magnetosonic-t mode\footnote{This dispersion relation for the magnetosonic-t mode is an approximation valid only in the strong magnetic field limit. Leading order corrections push $\omega$ to values slightly below $k_\gamma$, revealing the subluminal nature of this mode (see \eg~\cite{Gedalin1998}).}
\begin{equation}
    \omega^2 = k_\gamma^2 \, ,
\end{equation}
where $\theta_{kB}$ is the angle between $\vec{k}_\gamma$ and $\vec{B}_0$\footnote{Note that in our coordinates $\theta_{kB} = \theta_B$ at the resonance, but this need not be the case in general.}. For completeness, we note that in the case of propagation along the $y$-axis, the corresponding polarization vectors are given by
\begin{gather}
    \hat{\varepsilon}_{\rm LO} = \frac{1}{\sqrt{1 + \frac{\omega_p^4 \sin^2 \theta_{kB} \cos^2 \theta_{kB}}{(\omega^2 - \omega_p^2 \cos^2 \theta_{kB})^2}}} \Bigg(\hat{x} + \frac{\omega_p^2 \sin \theta_{kB} \cos \theta_{kB}}{\omega^2 - \omega_p^2 \cos^2 \theta_{kB}} \hat{y}\Bigg) \, , \\
    \hat{\varepsilon}_{\rm A} = \frac{-1}{\sqrt{1 + \frac{\omega_p^4 \sin^2 \theta_{kB} \cos^2 \theta_{kB}}{(\omega^2 - \omega_p^2 \cos^2 \theta_{kB})^2}}} \Bigg(\frac{\omega_p^2 \sin \theta_{kB} \cos \theta_{kB}}{\omega^2 - \omega_p^2 \cos^2 \theta_{kB}} \hat{x} - \hat{y}\Bigg) \, , \\
    \hat{\varepsilon}_{\rm mt} = \hat{z} \, .
\end{gather}
In this work we focus primarily on the superluminal LO mode, as it is the only one capable of escaping the plasma and generating an observable photon at Earth (the other modes instead represent locally confined excitations of the plasma), but comment briefly in later sections on the possibility of axion mixing with subluminous modes. 

The equations governing the evolution of the electric field in an anisotropic medium are given by
\begin{gather}
    \multlinegap=16pt
    \begin{multlined}
        \partial_y^2 E_x - \partial_x\partial_y E_y + (\omega^2 - \omega_p^2 \sin^2 \theta_B) E_x \\
        - \omega_p^2 \sin\theta_B \cos\theta_B E_y = - \omega^2 g_{a\gamma\gamma} B_0 \sin\theta_B a \, , \label{eq:coupled_aniso1}
    \end{multlined} \\
    \multlinegap=16pt
    \begin{multlined}\label{eq:coupled_aniso2}
        \partial_x^2 E_y - \partial_y\partial_x E_x + (\omega^2 - \omega_p^2 \cos^2 \theta_B) E_y \\
        - \omega_p^2 \sin\theta_B \cos\theta_B E_x = - \omega^2 g_{a\gamma\gamma} B_0 \cos\theta_B a \, ,
    \end{multlined}
\end{gather}
where we have again chosen our axions to travel in the positive $y$-direction (meaning our full system is contained within the $x$-$y$ plane). Notice that in the limit where $\theta_{B} \rightarrow \pi/2$, the first of the above equations coincides with its isotropic counterpart, which would yield the same conversion probability as in an isotropic medium. This result is not surprising, as this is the limit in which charged particles are freely allowed to oscillate in response to the perturbed electric field (without being affected by the constraint that they can move only along magnetic field lines)\footnote{This equivalence is, in principle, not exact. Photon refraction, induced \eg by variations in the plasma, will cause $\theta_B$ to evolve away from its value at resonance. For sufficiently slowly varying backgrounds, however, the equivalence holds.}. 

Solving Eqs.~\ref{eq:coupled_aniso1} and~\ref{eq:coupled_aniso2} is unfortunately not as trivial as in the previous cases. Various attempts have been made, each of which relies on a number of simplifying assumptions; we briefly review the two most recent approaches, and their results, here.

Ref.~\cite{millar2021axionphotonUPDATED} starts by arguing that the second order perpendicular derivative, $\partial_x^2$ in our setup, can be neglected in a slowly varying medium. This allows Eq.~\ref{eq:coupled_aniso2} to be rewritten in terms of the longitudinal electric field component, \ie
\begin{equation}\label{eq:Ey_full}
    E_y = \frac{\partial_y \partial_x E_x + \omega_p^2 \sin\theta_B \cos\theta_B E_x - \omega^2 g_{a\gamma\gamma} B_0 \cos\theta_B a}{\omega^2 - \omega_p^2 \cos^2\theta_B} \, ,
\end{equation}
which can subsequently be substituted into Eq.~\ref{eq:coupled_aniso1}. The authors then drop higher order derivatives (which again follows from the assumption of a slowly varying background), and adopt an Ansatz of the form
\begin{gather}
    E_x(x,y) = \tilde{E}_x(x,y) e^{i k_a y - i \omega t} \, , \\
    a(y) = a_0 e^{i k_a y - i \omega t} \, .
\end{gather}
Under these assumptions, Eq.~\ref{eq:coupled_aniso1} reduces to
\begin{multline}\label{eq:alex_oneeq}
    2 i k_a \partial_y \tilde{E}_x - 2 i k_a \frac{\omega_p^2 \xi}{\omega^2 \tan\theta_B} \partial_x \tilde{E}_x \\
    + (m_a^2 - \xi \omega_p^2 - ik \mathcal{D}) \tilde{E}_x \simeq - \frac{\omega^2 \xi}{\sin\theta_B} g_{a\gamma\gamma} B_0 a_0 \, ,
\end{multline}
where $\xi \equiv \sin^2\theta_B / (1- \omega_p^2 \cos^2\theta_B / \omega^2)$ and $\mathcal{D} \equiv 2 \omega_p \xi^2 / (\omega^2 \sin^2\theta_B \tan\theta_B) \partial_x \omega_p$. 

The authors finally define a new differential operator, $\partial_s \equiv \partial_y - \omega_p^2 \xi /(\omega^2 \tan\theta_B) \partial_x$, and use this to further reduce Eq.~\ref{eq:alex_oneeq} to a one-dimensional Schr\"{o}dinger-like equation. The solution can again be obtained by applying the stationary phase approximation about the resonant point, yielding an asymptotic value of $\tilde{E}_x$ equal to
\begin{equation}\label{eq:E_Alex}
    \tilde{E}_x \simeq \omega^2 g_{a\gamma\gamma} B_0 a_0 \, \sqrt{\frac{\pi}{2 k_a |\omega_p \partial_s \omega_p + \frac{\omega^2 - \omega_p^2}{\omega^2 \tan\theta_B} \omega_p^2 \partial_s \theta_B|}} \, .
\end{equation}
Here, an overall phase was disregarded, and the damping/growth term containing $\mathcal{D}$ has been argued to be small. As before, all quantities in Eq.~\ref{eq:E_Alex} are understood to be evaluated at the resonant point, which occurs when $k_a^\mu = k_\gamma^\mu$, or equivalently when
\begin{equation}\label{eq:resP}
    \omega_p^2 = \frac{m_a^2 \, \omega^2}{m_a^2 \cos^2\theta_B + \omega^2 \sin^2\theta_B} \, .    
\end{equation}
The final energy density stored via these excitations can be computed via
\begin{equation}
    U = \frac{1}{4}\partial_\omega (\omega \epsilon^{ij}) E_i^* E_j + \frac{1}{4}B^*_i B^j \, .
\end{equation}
Taking the ratio of this energy density with that of the initial axion field yields the following conversion probability~\cite{millar2021axionphotonUPDATED}\footnote{Note that there was an error in the original derivation of this conversion probability arising from an inconsistency in the non-relativistic limit. It was assumed that $m_a \simeq \omega_p$ at the resonance, which is only true at zeroth order in velocity, while the remaining equations were expanded to second order. This lead to a non-trivial error that has been corrected in the equations outlined here.}
\begin{multline}\label{eq:pa_alex}
    P_{a \rightarrow \gamma}^{\rm ani1} = \frac{\pi}{2} \left(1 + \frac{\omega_p^4 \xi^2}{\omega^4 \tan^2 \theta_B}\right) \frac{\omega^2 g_{a\gamma\gamma}^2 B_0^2}{k_a} \\
    \times \frac{1}{|\omega_p \partial_s \omega_p + \frac{\omega^2 - \omega_p^2}{\omega^2 \tan\theta_B} \omega_p^2 \partial_s \theta_B|} \, .
\end{multline}

An alternative method based on kinetic theory was adopted by~\cite{mcdonald2023axionphoton}. Here, the authors derive a Boltzmann-like transport equation for the photon distribution function. This describes the evolution of photons along worldlines, and includes a non-zero source term produced by the interaction with axions at a resonance. Assuming that the production/absorption of photons due to the plasma is negligible, this equation is given by
\begin{equation}\label{eq:jamie_boltz}
    \partial_k \mathcal{H} \partial_x f_\gamma - \partial_x \mathcal{H} \partial_k f_\gamma = \varepsilon^*_\mu \varepsilon_\nu \Pi_{\rm ax}^{< \mu\nu} \, .
\end{equation}
Eq.~\ref{eq:jamie_boltz} considers the production of a particular electromagnetic mode, which has distribution function $f_\gamma$, Hamiltonian $\mathcal{H}(x,k)$ (defined such that $\mathcal{H} = 0$ gives the dispersion relation for the mode), and polarization four-vector $\varepsilon_\mu$. Moreover, $\Pi_{\rm ax}^{< \mu\nu}$ is the Wigner transform of the photon Wightman self-energy, which describes photon production from axions. The latter is defined as
\begin{equation}
    \Pi_{\rm ax}^{< \mu\nu}(x, x^\prime) = \left< j^\nu_a(x^\prime) j_a^\mu(x) \right> \, ,
\end{equation}
where the axion-photon current has the form $j^\mu_a = g_{a\gamma\gamma} \tilde{F}^{\mu\nu}_{\rm ext} \partial_\nu a$, with $\tilde{F}^{\mu\nu}_{\rm ext}$ being the dual external field strength tensor (which in our case carries the contribution from the magnetic field of the neutron star). After the Wigner transform, and after performing an expansion of $\tilde{F}_{\rm ext}$ about the resonant point, the authors arrive at the following expression
\begin{equation}
    \Pi_{\rm ax}^{< \mu\nu} = g_{a\gamma\gamma}^2 k_\rho k_\sigma \tilde{F}^{\mu\rho}_{\rm ext} \tilde{F}^{\nu\sigma}_{\rm ext} 2\pi \delta(k^2 - m_a^2) f_a \, ,
\end{equation}
where $f_a$ is the axion distribution function. 

By applying the method of characteristics, which identifies the photon world lines $(x^\mu(\lambda), k_\mu(\lambda))$, Eq.~\ref{eq:jamie_boltz} can be rewritten in terms of the evolution of the photon distribution function
\begin{equation}
    \frac{df_\gamma(x(\lambda), k(\lambda))}{d\lambda} = g_{a\gamma\gamma}^2 |k \cdot \tilde{F}_{\rm ext} \cdot \varepsilon|^2 2\pi \delta(\omega_\gamma(\lambda)^2 - \omega_a(\lambda)^2) f_a \, .
\end{equation}
Integrating this equation, and defining $P_{a\rightarrow \gamma} = f_\gamma / f_a$, finally yields the following expression for the conversion probability~\cite{mcdonald2023axionphoton}
\begin{equation}\label{eq:pa_jamie}
    P_{a \rightarrow \gamma}^{\rm ani2} = \frac{\pi}{2} \frac{\omega^4 g_{a\gamma\gamma}^2 B_0^2 \sin^2\theta_B}{\omega_p^2(\omega_p^2 - 2 \omega^2) \cos^2\theta_B + \omega^4} \frac{1}{|\vec{v}_p \cdot \nabla \omega |} \, ,
\end{equation}
where $\vec{v}_p \equiv k / \omega$ denotes the photon phase velocity and all parameters are again set to their value at the resonant point. We note that this conversion probability and Eq.~\ref{eq:pa_alex} can be more explicitly compared if one evaluates the gradient of the energy. An important distinction worth bearing in mind is that~\cite{mcdonald2023axionphoton} adopts a Hamiltonian formalism in which space and momentum are conjugate variables, whereas the calculation performed in~\cite{millar2021axionphotonUPDATED} (and the calculations we have provided above for free space and an isotropic plasma) assume instead that the energy is fixed. The gradient of the energy in this case takes the form
\begin{equation}
    \nabla \omega = \frac{\omega^3 \omega_p \sin^2\theta_B \nabla \omega_p + \omega \omega_p^2 \sin\theta_B \cos\theta_B(\omega^2 - \omega_p^2) \nabla \theta_B}{\omega_p^2(\omega_p^2 - 2\omega^2)\cos^2\theta_B + \omega^4} \, .
\end{equation}
In our notation, the phase velocity at the resonant point is given by $\vec{v}_p = (0, k_a / \omega, 0)$, and thus the conversion probability can be rewritten as
\begin{equation}\label{eq:pani2}
    P_{a \rightarrow \gamma}^{\rm ani2} = \frac{\pi}{2} \frac{\omega^2 g_{a\gamma\gamma}^2 B_0^2}{k_a} \frac{1}{|\omega_p \partial_y \omega_p + \frac{\omega^2 - \omega_p^2}{\omega^2 \tan\theta_B} \omega_p^2 \partial_y \theta_B|} \, .
\end{equation}
It is now clear that this differs only from Eq.~\ref{eq:pa_alex} in terms of the pre-factor in large brackets, and the direction with respect to which the derivatives are computed.

A few comments are in order at this point. First, in the numerical problem constructed in this work, $\partial_y \theta_B = 0$. Naively, this seems to imply that Eq.~\ref{eq:pani2} does not depend on $\theta_B$; this conclusion, however, is not correct, as all quantities in Eq.~\ref{eq:pani2} are evaluated at the resonance, which itself has an implicit dependence on $\theta_B$, see Eq.~\ref{eq:resP} (although the dependence does drop out in the non-relativistic limit). Second, in the limit where $\theta_B \rightarrow 0$, the resonance point in Eq.~\ref{eq:resP} goes to $\omega \simeq \omega_p$. This is a rather peculiar point, where all three dispersion relations are degenerate, potentially allowing for mode mixing (note that this would invalidate the analytic derivations outlined above). We return to this peculiar point in the following section. 

Despite the fact that Eqs.~\ref{eq:pa_alex} and~\ref{eq:pa_jamie} seem to be based on the same set of fundamental assumptions (namely, that the background varies slowly with respect to the axion/photon wavelength, that the axion is treated perturbatively to leading order, and that the axion is assumed to mix exclusively with the LO mode), the results can differ markedly (see Fig.~\ref{fig:pa_th_ph_v_0.4_0.8} in the following section). In this paper, we attempt to resolve this discrepancy by directly solving for the solution in a regime where all underlying assumptions are expected to hold; in future work we also intend to break these assumptions, and instead investigate the behavior of axion-photon mixing in more extreme scenarios.

\subsection{Numerical approach}
Having outlined a number of analytic approximations which characterize the efficiency of axion-photon mixing in various settings, we now turn our attention toward the numerical approach developed in this paper, which directly solves Eq.~\ref{eq:waveeq} (or equivalently Eqns.~\ref{eq:coupled_aniso1} and~\ref{eq:coupled_aniso2}) for fixed assumptions of the dielectric tensor and the background magnetic field. At this point it is important to emphasize that our goal is not to precisely replicate the physical problems of interest, but rather to test the precision of analytic expressions, and illustrate the power of numerical calculations in addressing axion-photon mixing in arbitrary systems. Therefore, in order to ease our calculations, we construct problems that resemble the case of axion mixing near neutron stars, but with a reduced scale separation. In future work we intend to increase the complexity of this framework, slowly working toward a reconstruction of axion-photon mixing using the physical scales and parameters that are appropriate for extreme astrophysical environments.


\begin{figure}[ht!]
    \includegraphics[width=0.46\textwidth,trim={0.0cm 0.0cm 0.0cm 0cm}, clip] {"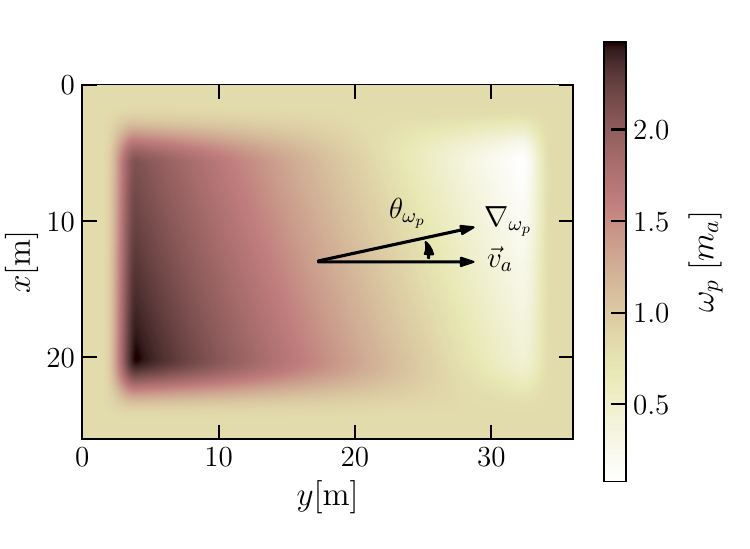"}
    \includegraphics[width=0.46\textwidth, trim={0cm 0cm 0cm 0cm}, clip] {"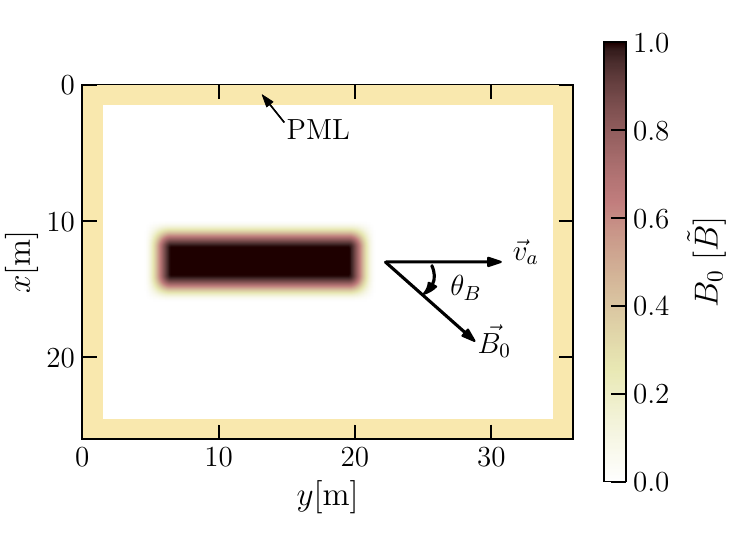"}
    \includegraphics[width=0.4\textwidth, trim={0cm 0cm 0cm 0cm}, clip] {"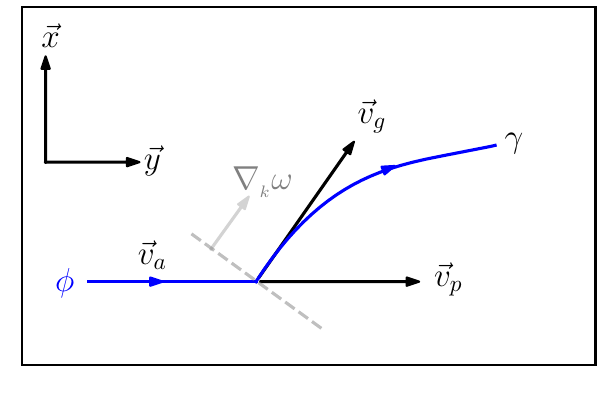"}			
	\caption{{\it Top:} An example plasma frequency profile within our two-dimensional simulation domain. The plasma gradient angle is set to $\theta_{\omega_p} = 20^\circ$ and highlighted in the plot. The units of $\omega_p$ are normalized to the axion mass $m_a$. {\it Middle:} Our fiducial magnetic field profile, shown together with the PML. The definition of the magnetic field angle $\theta_B$ is also illustrated. Here, we have normalized $B_0$ to a value $\tilde{B}$, which we fix throughout this work to be $\tilde{B} = 1.95 \times 10^{-15} \, \mathrm{GeV}^{-2}$ (the precise value being unimportant for the details of this work). {\it Bottom:} Depiction of the axion and photon velocities for an example trajectory. The photon phase and group velocity are denoted as $\vec{v}_p$ and $\vec{v}_g = \nabla_k \omega$ respectively. }
    \label{fig:simdetails}
\end{figure}

\begin{table*}[t]
    \centering
    \begin{tabular}{c|c|c}
        \hline\hline
        \textbf{Name} & \textbf{Fiducial value} & \textbf{Definition}  \\
        \hline
        $L_x$ & $26 \, \rm m$ & Width of the simulation domain \\
        $L_y$ & $36 \, \rm m$ & Length of the simulation domain \\
        $L_{B_x}$ & $2 \, \rm m$ & Width of the magnetic field region \\
        $L_{B_y}$ & $13 \, \rm m$ & Length of the magnetic field region \\
        $\delta L_B$ & $2 \, \rm m$ & Size of the $\sin^2$ smoothing region of the magnetic field \\
        $B_0$ & $10 \, \rm T$ & Magnetic field strength \\
        $\theta_B$ & -- & Clockwise angle between the axion momentum and the magnetic field \\
        $\theta_{\omega_p}$ & -- & Counterclockwise angle between the axion momentum and the plasma gradient \\
        $\alpha$ & -- & Slope of the plasma frequency profile \\
        $\beta$ & -- & Constant term in the plasma frequency profile \\
        $a_0$ & $2.63 \times 10^{-8} \, \mathrm{GeV}$ & Axion field normalization \\
        $m_a$ & $10^{-5} \, \rm eV$ & Axion mass \\
        $g_{a\gamma\gamma}$ & $2 \times 10^{-14} \, \mathrm{GeV}^{-1}$ & Axion-photon coupling \\
        $v_a$ & -- & Axion velocity \\
        \hline\hline
    \end{tabular}
    \caption{Simulation parameters together with their fiducial value and definition. }
    \label{tab:simparam}
\end{table*}

For simplicity, we confine our analysis to two spatial dimensions, assuming that the axion, the plasma gradient, and the magnetic field all lie in the same plane. We adopt a fiducial simulation domain size of $26 \times 36$ meters (corresponding to the length of the simulation domain in the $x$- and $y$-directions respectively). A uniform magnetic field is placed within a subset of this region, spanning a distance of $2 \times 13$ meters, as shown in the center pane of Fig.~\ref{fig:simdetails}. In order to avoid spurious numerical effects arising from sharp boundaries of the magnetic field, we introduce a $\sin^2$ smoothing boundary such that the magnetic field (and its gradient) smoothly drop to zero -- the size of this boundary is parameterized by $\delta L_B$ (in both dimensions). Using coordinates centered on the magnetic field region, this implies \eg that the magnetic field in the first quadrant scales as $|\vec{B}| \propto \sin^2(\frac{\pi}{2} (\frac{y}{\delta L_B} + 1 - \frac{L_{B_y}}{2 \delta L_B})) \cdot \sin^2(\frac{\pi}{2} (\frac{x}{\delta L_B} + 1 - \frac{L_{B_x}}{2 \delta L_B}))$, with similar expressions being applied in the other three quadrants\footnote{Note that our magnetic field does not satisfy Gauss' law of magnetism, $\nabla \cdot \vec{B} = 0$. In principle, this can induce radiative corrections to the induced electric fields; these corrections, however, should remain small so long as the gradient is small with respect to the axion wavelength (see \eg~\cite{knirck2019first}).}. The magnetic field orientation is parameterized by an angle $\theta_B$, which is defined with respect to the $y$-axis and rotated in the direction of the $x$-axis. A perfectly matched layer (PML), acting as an artificial absorbing layer, is constructed around the boundary of the domain and helps to avoid spurious boundary effects that can arise when applying finite element method solvers to electromagnetism. We utilize the uniaxial formulation of PMLs, which amounts to re-scaling the dielectric tensor by a uniaxial anisotropic tensor. As explained in~\cite{taf05}, this tensor is parameterized by one complex parameter for each dimension, with $\kappa$ and $\sigma$ defining the real and imaginary components respectively. In this work we have used a polynomial grading approach, which implements the following profiles on each dimension: $\sigma_x(x) = (x/d)^m \sigma_{x, \rm max}$ and $\kappa_x(x) = 1 + (x/d)^m(\kappa_{x, \rm max} - 1)$ (and similarly for the $y$-coordinate). In these expressions, $d$ is the width of the PML (which we set to $1.5$ meters), $m$ is an order one free parameter which we fix to $m = 3$, and $\sigma_{x/y, \rm max}$ has been varied to optimize the efficiency of the PML. 

We also incorporate a spatially varying plasma frequency over the simulation domain. Throughout most of the domain, we take a linearly varying plasma profile 
$\omega_p(\ell) = \alpha \,  \ell + \beta$, where the direction of $\ell$ (\ie the plasma gradient) is chosen to lie in the $x$-$y$ plane, and sits at an angle $\theta_{\omega_p}$ as defined with respect to the $y$-axis (see Fig.~\ref{fig:simdetails}); we vary $\theta_{\omega_p}$ across simulations in order to explore the scaling of the conversion probability. The free parameters $\alpha$ and $\beta$ are chosen in each simulation such that the resonance occurs at the center of the magnetic field region. In order for the PML to serve as an effective absorbing layer, the dielectric tensor must be homogeneous near the edge of the box. Consequently, we introduce a smoothing region that re-scales the plasma profile by the same $\sin^2$ profile that was used to smooth the magnetic field region (here, we take the length of the smoothing region to be $2.5$ meters and $4.5$ meters in the $x$- and $y$-directions respectively) -- this transitions the plasma profile from a linear scaling to zero near the boundary. The full simulation set-up can be seen in Fig.~\ref{fig:simdetails}; the parameters entering the simulation, as well as their fiducial values, can be found in Table~\ref{tab:simparam}. We have verified that alternative choices for \eg the simulation domain size and the magnetic field region do not change our results.

We adopt a fiducial nodal separation of $7.5 \, \rm mm$, corresponding to a total of $\sim 10$ million nodes in the two-dimensional simulation box. This number is chosen to ensure that all calculations are performed with a minimum of $10$ nodes per wavelength. Solutions are reconstructed using the finite element solver provided by the Elmer FEM simulation software~\cite{ElmerRef} (note that we adopt a direct, rather than an iterative, solver, as the direct solver resolves the large sparse matrices exactly, implying it is not affected by convergence issues -- although this comes at the expense of larger memory consumption). In the examples that follow, we adopt a fiducial axion mass of $10^{-5} \, \rm eV$, and consider two choices of the axion velocity, corresponding to $v_a = 0.4$ and $v_a = 0.8$ (note that taking $v_a \rightarrow 0$ eventually becomes problematic, as the axion wavelength becomes much large than the resonance region). In the context of axion-photon conversion near neutron stars, axions are typically close to the semi-relativistic regime, with $v_a \sim 0.5$, and thus our choices capture the typical scales that arise in realistic scenarios. 

The relevant observable that we extract from each simulated run is the axion-photon conversion probability $P_{a \rightarrow \gamma}$, which is simply the probability that an incoming axion with momentum $k^\mu_a = (\omega, \vec{k}_a)$ converts to a photon after crossing a resonance located at $k^\mu_a = k^\mu_\gamma$. In order to make contact with analytic descriptions of this process, we note that
the conversion probability is related to the rate of flow of electromagnetic energy $\mathcal{S}$ through a surface $dA$ via~\cite{mcdonald2023axionphoton,mcdonald2023generalized}
\begin{equation}
   \mathcal{S} \equiv \int d^3 k \int d A \cdot \vec{v}_g \, \omega f_\gamma = \int d^3 k \int d \Sigma_k \cdot \vec{v}_a \, \omega P_{a \rightarrow \gamma} f_a \, ,
\end{equation}
where $f_a$ is the axion phase space function, $\vec{v}_a$ is the axion velocity, $\vec{v}_g$ is the group velocity of the photon, $\Sigma_k$ is the surface defined by the set of points over which $k^\mu_a = k^\mu_\gamma$, and $dA$ is a bounding surface around the resonance.

In the context of the simulation, this implies that the conversion probability is related to the integration of the Poynting flux $\vec{S}$ over the edge of the simulation boundary (defined here to be the region just inside the PML), via
\begin{equation} \label{eq:pa_fluxes}
    P_{a \rightarrow \gamma} = \frac{2 \int dA \cdot \vec{S}}{a^2 \omega k_a L_{B_x} \zeta(\theta_{\omega_p})} \, .
\end{equation}
Here, $\zeta(\theta_{\omega_p})$ is a function that describes the fraction of the axion flux which crosses the conversion surface. For our fiducial setup, this function is $\zeta = 1$ for $\theta_{\omega_p} \lesssim 81^\circ$, but goes to zero in the limit that $\theta_{\omega_p} \rightarrow 90^\circ$ (corresponding to the scenario in which the conversion surface is perpendicular to $\vec{k}_a$). In order to avoid having to compute this function for large angles of incidence, we instead choose to elongate the magnetic field in the $y$-direction such that $\zeta$ is always one for all values of $\theta_{\omega_p}$. Importantly, Eq.~\ref{eq:pa_fluxes} neglects conversion taking place in the boundary region of the magnetic field, where the magnetic field strength is suppressed by $\sin^2$. We incorporate this correction by defining an effective $L_{B_x}^{\rm eff}$ that accounts for the total incoming integrated power flow across both the constant and the smoothed $\sin^2$ regions. 

While the primary observable is the conversion probability, we will illustrate in the following sections that the numerical approach adopted here also has the ability to disentangle the excitation of the three distinct electromagnetic modes that are present in a highly magnetized plasma. In order to illustrate this, we need to extract the momentum $\vec{k}_\gamma$ from each point in the simulation box, and show that this quantity, when combined with the background plasma frequency, can be accurately used to reconstruct each dispersion relation. In order to do so, we compute the phase $\psi$ of the electric field by using the following relation
\begin{eqnarray}\label{eq:wavevector}
    \vec{k}_\gamma = \nabla \psi &=& \nabla \left(\frac{\mathrm{ln} E_x - \mathrm{ln} E_x^*}{2 i}\right) = \frac{1}{2i}\left(\frac{\nabla E_x}{E_x} - \frac{\nabla E_x^*}{E_x^*}\right) \nonumber \\ 
    &=& \frac{1}{2i}\left(\frac{E_x^*\nabla E_x-E_x \nabla E_x^*}{E_x E_x^*}\right) = \frac{E_x \overleftrightarrow{\nabla}E_x^*} {2 i E_x E_x^*} \, .
\end{eqnarray}
The momentum can also be obtained with an equivalent expression featuring the $E_y$ component, yielding identical results.

\section{Results}\label{sec:results}
We present here the results of our numerical analysis, including a critical comparison to the known analytic formulas derived in Sec.~\ref{sec:method}. We begin with the simplified scenarios in which the analytic expressions for the conversion probability are recognized to be under control; these include the case of non-resonant mixing in vacuum, and the case of resonant mixing in the isotropic limit. After showing the high level of agreement that can be obtained here, we turn our attention to the more complex scenario of resonant mixing in an anisotropic background. 

\begin{figure}
	\centering
        \includegraphics[width=\linewidth] {"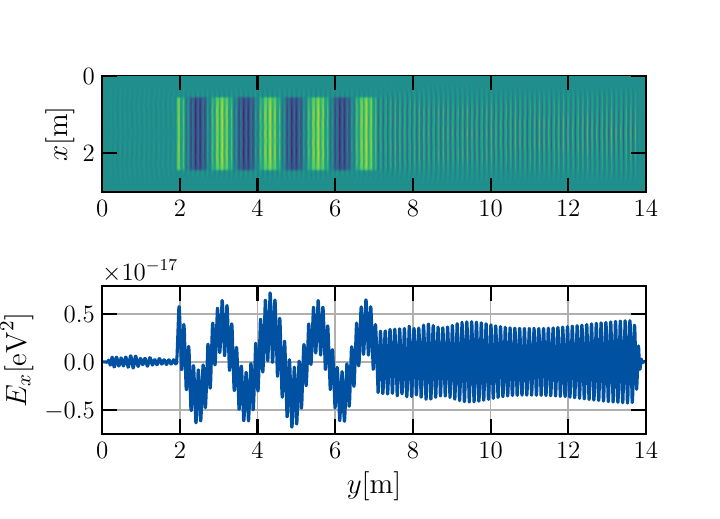"}
        \caption{{\it Top:} The amplitude of $E_x$ for an axion traversing a magnetic field in free space. {\it Bottom:} The evolution of $E_x$ in a one-dimensional slice through the center of the box. For both plots the axion velocity is set to $v_a = 0.1$, the magnetic field is chosen to point in the $x$-direction, and all other parameters (except for the box and magnetic field region size) are set to their fiducial values. }
	\label{fig:Ex_freespace}
\end{figure}

\subsection{Free space}
In vacuum, axions can only excite on-shell photons with the vacuum dispersion relation $\omega = k_\gamma$. However, inside the magnetic field region axions will also induce an oscillatory electric field with $\omega = k_a$. The appearance of both modes can be clearly distinguished in Fig.~\ref{fig:Ex_freespace}, which depicts $E_x$ across the simulation domain (note that this figure is made using a simulation domain that is smaller than our fiducial domain detailed in Sec.~\ref{sec:method} -- this is done to ensure both modes can be distinguished without the need for a zoom-in). Here, one can see both large amplitude long-wavelength fluctuations, corresponding to the induced axion electric field, and small amplitude short-wavelength oscillations, arising from the newly sourced on-shell photon. As expected, only the on-shell photon persists in the limit $y \rightarrow \infty$. 

\begin{figure}
	\centering
        \includegraphics[width=\linewidth] {"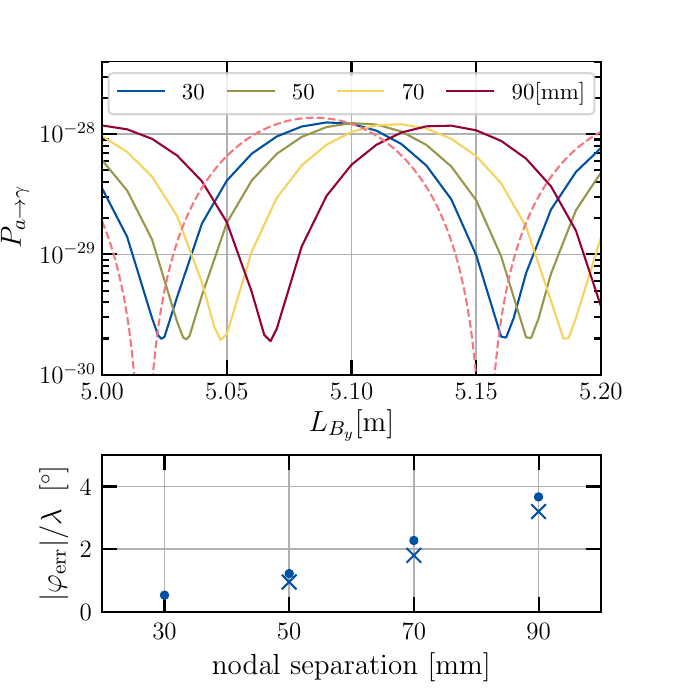"}
        \caption{{\it Top:} Numerical (solid) and analytic (dashed) conversion probability in free space as a function of the length of the background magnetic field region in the $y$-direction, $L_{B_y}$. Four curves are computed using different nodal separations in order to illustrate the effect of the phase-shift which arises from solving the equations at finite resolution. The analytic conversion probability is calculated with Eq.~\ref{eq:P_free}, and all parameters are set to the same values as used in Fig.~\ref{fig:Ex_freespace}. {\it Bottom:} The phase shift per wavelength (in degrees) from our simulation results (dots) for nodal separations of $50$, $70$ and $90 \, \rm mm$ is compared to the theoretical results (crosses) for a plane wave propagating through a mesh of bilinear elements at $0$ degrees, as computed in Fig.~4 of~\cite{Lee:1992} (note that Ref.~\cite{Lee:1992} does not contain results for $30 \, \rm mm$, and thus no cross is shown here). }
	\label{fig:pa_freespace}
\end{figure}

\begin{figure}
	\centering
        \includegraphics[width=0.85\linewidth, trim={.1cm .4cm .5cm .5cm}, clip] {"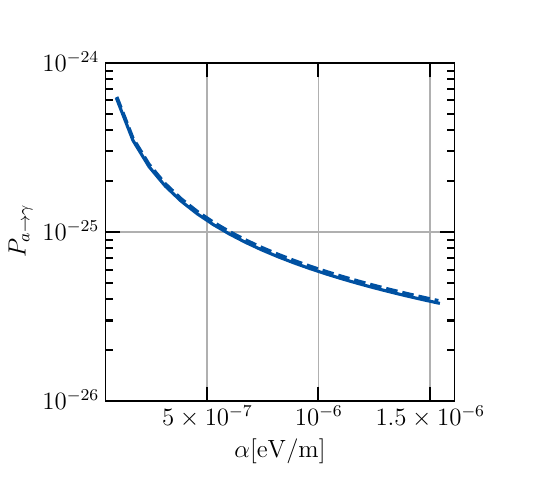"}	  \includegraphics[width=0.85\linewidth, trim={.1cm .4cm .5cm .5cm}, clip] {"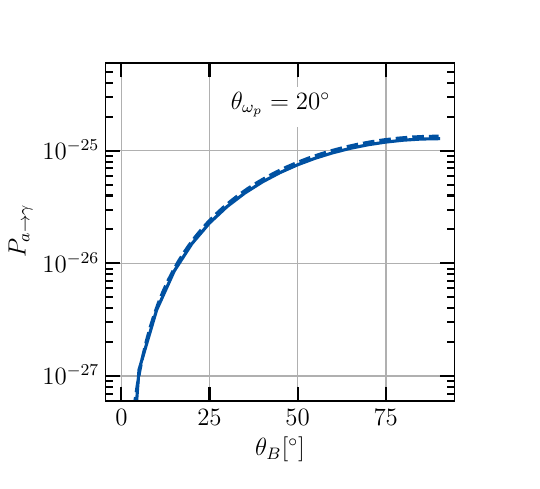"}				        \includegraphics[width=0.85\linewidth, trim={.1cm .4cm .5cm .5cm}, clip] {"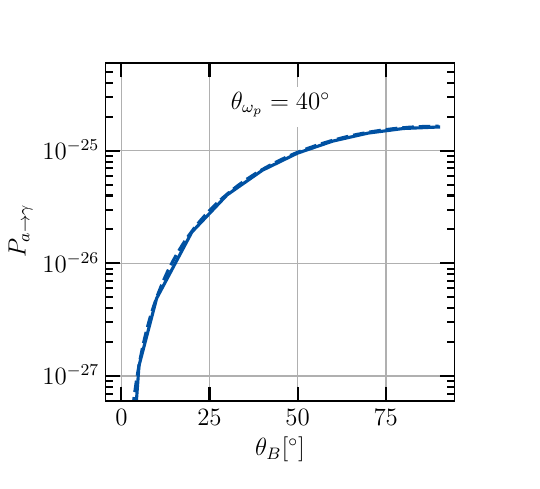"}				
        \caption{{\it Top:} Numerical (solid) and analytic (dashed) conversion probability in an isotropic plasma as a function of the gradient of the plasma frequency, $\alpha$. Here we have set $\theta_B = 90^\circ$ and $\theta_{\omega_p} = 0^\circ$. {\it Center, Bottom:} Numerical (solid) and analytic (dashed) conversion probability in an isotropic plasma as a function of $\theta_B$, setting $\theta_{\omega_p}=20^\circ$ (center) and $\theta_{\omega_p}=40^\circ$ (bottom). Note the scale on the y-axis should not be interpreted as being representative of any particular problem, and can be freely re-scaled to any system of interest. }
	\label{fig:pa_alpha}
\end{figure}

The conversion probability of axions to on-shell photons in free space, computed using Eq.~\ref{eq:pa_fluxes}, is plotted in Fig.~\ref{fig:pa_freespace} as a function of the length of the magnetic field domain in the $y$-direction. The results are shown for four different choices of resolution, corresponding to nodal separations ranging between $30 \, \rm mm$ and $90 \, \rm mm$, and compared with the analytic calculation, Eq.~\ref{eq:P_free}. Up to a phase shift, the numerical results are in excellent agreement with the analytic derivation. This phase shift is a well-known effect that arises from spatial discretization, with the value of the shift $\varphi_{\mathrm{err}}$ (in radians) accumulated over a length $l$ being set by the difference in the physical ($k_\gamma$) and numerical ($\hat{k}_\gamma$) wave numbers~\cite{Lee:1992}, \ie
\begin{equation}\label{eq:discretization_error}
    \varphi_{\mathrm{err}}=(k_\gamma - \hat{k}_\gamma) l \, .
\end{equation}
 Fig.~4 of~\cite{Lee:1992} shows the calculated phase error per wavelength as a function of the nodal density per wavelength. We compare these results with the phase shift of our numerical results in the bottom panel of Fig.~\ref{fig:pa_freespace}. 

This phase shift, and the subsequent sensitivity of the inferred conversion probability to the spatial resolution, arise because photon production is sensitive to the Fourier transform of the magnetic field, which itself is an oscillatory function of $L_{B_y}$. Consequently, this is only a feature of non-resonant mixing, and does not arise for the resonant scenarios outlined below.

\begin{figure*}
	\centering
        \includegraphics[width=\linewidth] {"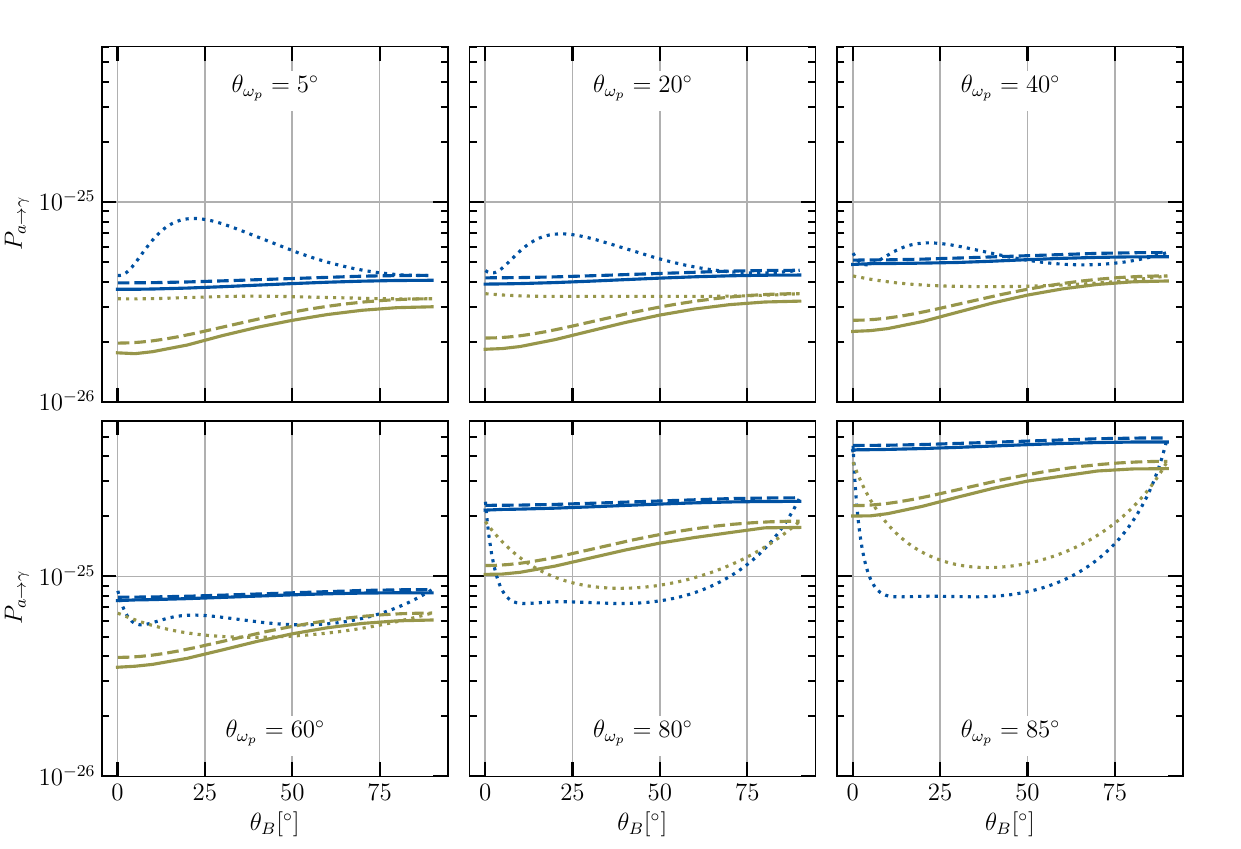"}			
	    \caption{Numerical (solid) and analytic (Eqs.~\ref{eq:pa_alex} and~\ref{eq:pa_jamie} as respectively dotted and dashed) conversion probability in an anisotropic plasma as a function of $\theta_B$ and $\theta_{\omega_p}$. We plot the results for both $v_a = 0.4$ (blue) and $v_a = 0.8$ (yellow). Note the scale on the $y$-axis should not be interpreted as being representative of any particular problem, and can be freely re-scaled to any system of interest. This plot is the main result of the paper. Our numerical results show excellent agreement with the analytic derivation from~\cite{mcdonald2023axionphoton}. As described in the main text and the Appendix, the small observed offset between our results and those of~\cite{mcdonald2023axionphoton} arises from limited nodal resolution. }
	\label{fig:pa_th_ph_v_0.4_0.8}
\end{figure*}

\subsection{Isotropic plasma}
We now turn our attention to resonant axion-photon mixing in an isotropic plasma. In Fig.~\ref{fig:pa_alpha}, we plot the conversion probability as a function of the plasma gradient which is controlled by the parameter $\alpha$ (top), and as a function of $\theta_B$ for two fixed example values of $\theta_{\omega_p}$ (center and bottom). The top plot considers the simple case in which the axion is traveling in the direction parallel to the gradient of the plasma frequency, and in which the magnetic field is oriented in the perpendicular direction. The results are  in excellent agreement with the analytic conversion probability given in Eq.~\ref{eq:prob_iso}, which provides strong confidence that our numerical scheme is capable of resolving the dynamics of resonances in these systems\footnote{As a word of caution to those that follow in our footsteps: the axion-induced electric field (\ie not the propagating mode) can be resonantly excited when $\omega \simeq \omega_p$, and this resonance can significantly disrupt the ability to numerically resolve the sourced on-shell mode. In order to avoid this we have constructed the plasma profile in such a way that this secondary resonance always falls outside the magnetic field region.}. 

\subsection{Anisotropic plasma}
Finally, we turn our attention to the scenario of interest, resonant axion-photon mixing in an anisotropic plasma. One of the main results of this paper is Fig.~\ref{fig:pa_th_ph_v_0.4_0.8}, which displays the inferred value of the conversion probability as a function of $\theta_B$ and $\theta_{\omega_p}$ for two choices of the axion velocity ($v_a = 0.4$ and $v_a = 0.8$). For comparison, we plot alongside our results the analytic conversion probabilities from Eqs.~\ref{eq:pa_alex} and~\ref{eq:pa_jamie}. Fig.~\ref{fig:pa_th_ph_v_0.4_0.8} shows strong agreement with the latter across all of the parameter space (while we acknowledge a slight underestimation is visible, we show in the Appendix via convergence tests that this is likely due to issues of numerical resolution). Our results therefore seem to resolve the discrepancy between~\cite{mcdonald2023axionphoton} and~\cite{millar2021axionphotonUPDATED}, and validate the use of Eq.~\ref{eq:pa_jamie} for resonant axion-photon mixing near neutron stars (at least in the regime where the WKB approximation is assumed to remain valid -- as mentioned above, this approximation is not expected to hold \eg near the open field lines of neutron stars, or for very low-energy axions).

\begin{figure*}
	\centering
    \includegraphics[width=\linewidth] {"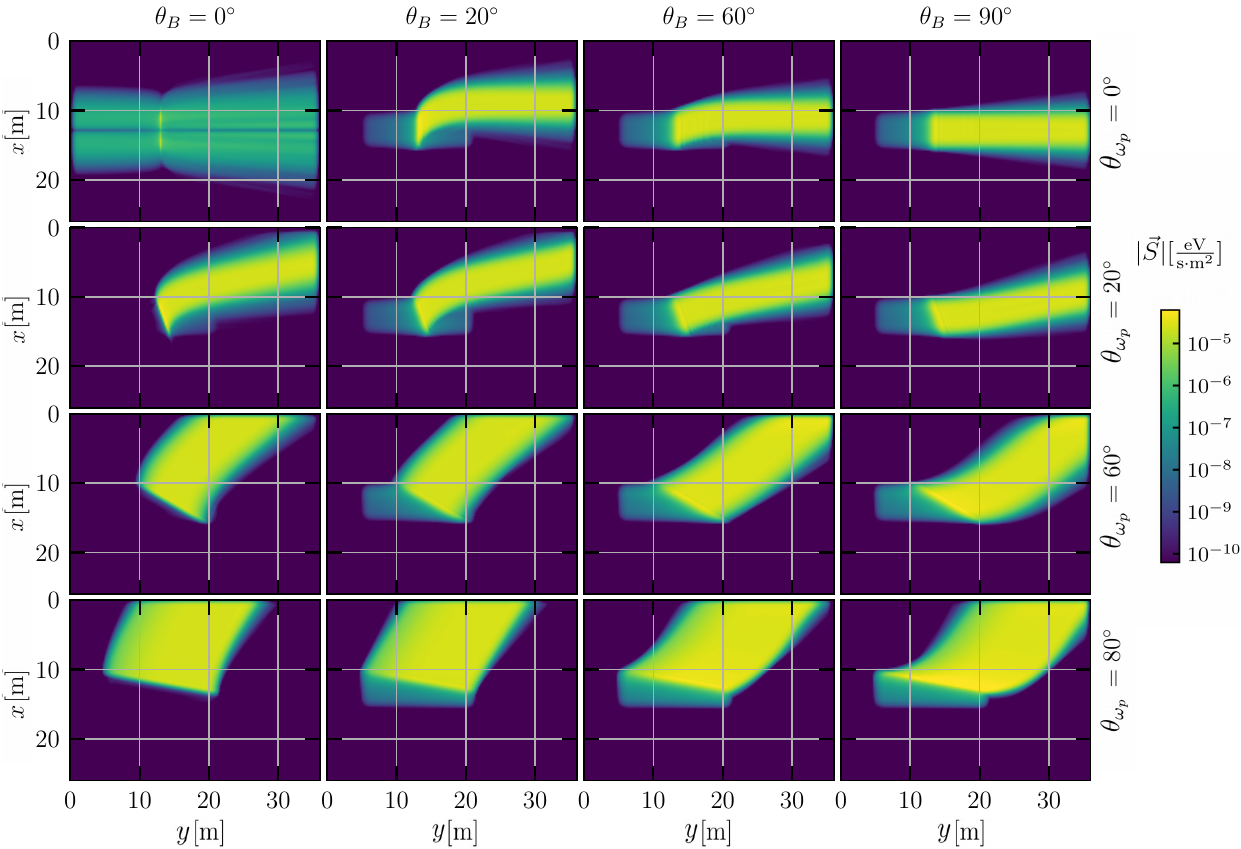"}	
	\includegraphics[width=\linewidth] {"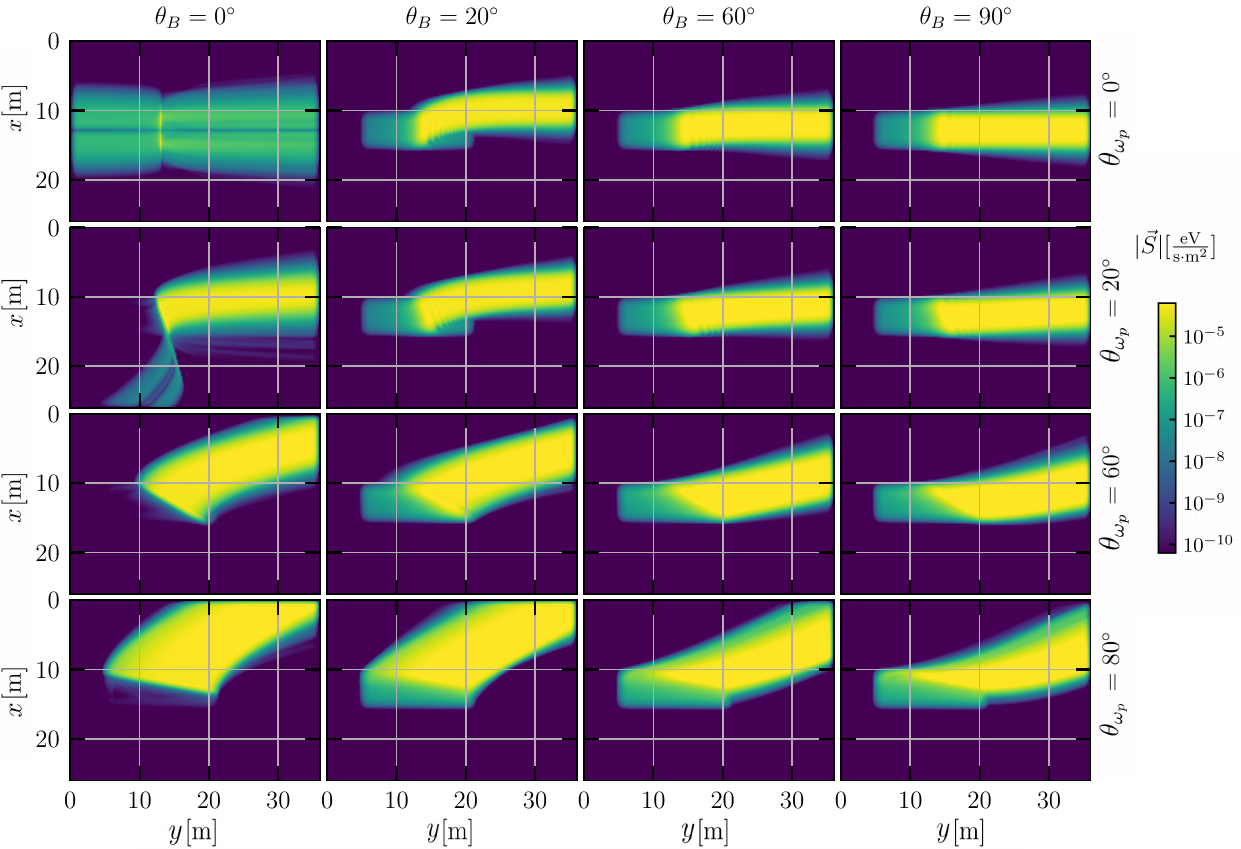"}	
	\caption{{\it Top:} Magnitude of the Poynting flux for $v_a = 0.4$. {\it Bottom:} The same for $v_a = 0.8$. We plot for various choices of $\theta_{B}$ and $\theta_{\omega_p}$. }
	\label{fig:S_th_ph}
\end{figure*}

We also show in Fig.~\ref{fig:S_th_ph} the magnitude of the Poynting flux for the case of $v_a = 0.4$ (top) and $v_a = 0.8$ (bottom). There are two interesting features which emerge from this figure. First, one can differentiate the effect of refraction, which in this example is induced purely by the direction of the plasma gradient, from the misalignment of the group and phase velocities of the electromagnetic waves (the former lying in the $x$-$y$ plane, and the latter pointing in the $y$-direction). We also illustrate this point more clearly using streamline plots in the Appendix. Next, one can see that there is a peculiar behavior at $\theta_B = \theta_{\omega_p} = 0$. This is the point in parameter space where the resonance occurs when all modes are degenerate, as are the group velocities of the LO and Alfv\'{e}n modes. This figure provides an indication that non-trivial mixing may occur in certain, albeit somewhat tuned, circumstances. We leave a deeper investigation of axion-induced mode mixing to future work.

The conversion probabilities computed in Fig.~\ref{fig:pa_th_ph_v_0.4_0.8} reflect the on-shell production of LO modes, which has been the primary focus for axion studies in magnetized plasma due to the fact that such photons can escape the magnetosphere and produce observable radio emission. In general, axions can couple to other modes, and in certain cases different electromagnetic modes can themselves mix. Whether such effects can ever be relevant in physical systems remains an open question. Given that our approach naturally incorporates both of these possibilities, we take the opportunity to further analyze our simulations and explore the extent to which either of these effects arise.

\begin{figure*}
	\centering
        \includegraphics[width=0.8\linewidth] {"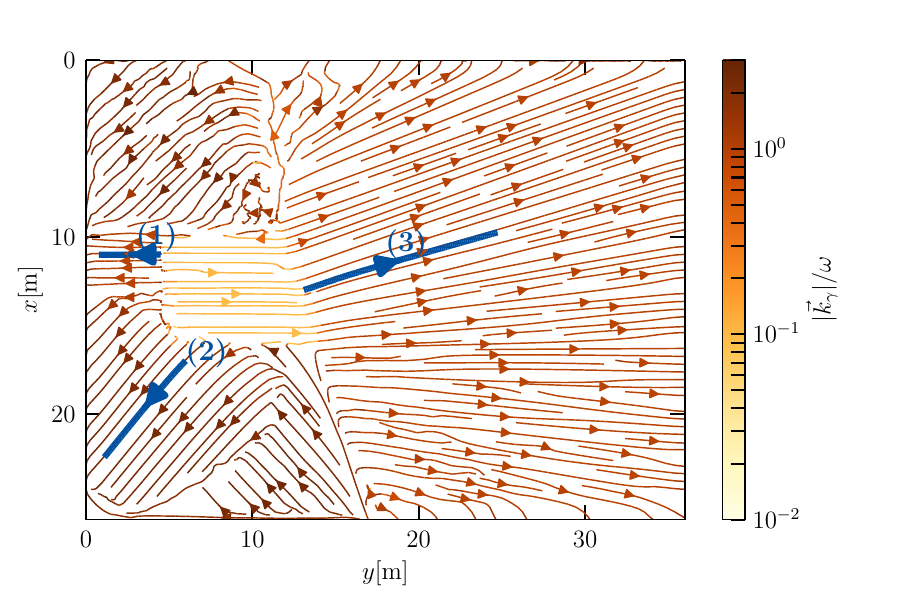"}			
        \includegraphics[width=0.95\linewidth] {"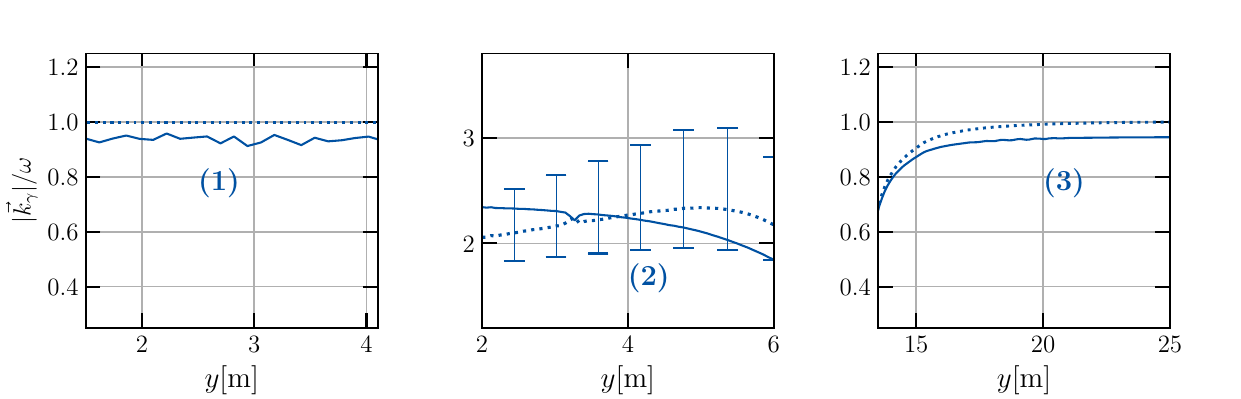"}	
        \includegraphics[width=0.95\linewidth] {"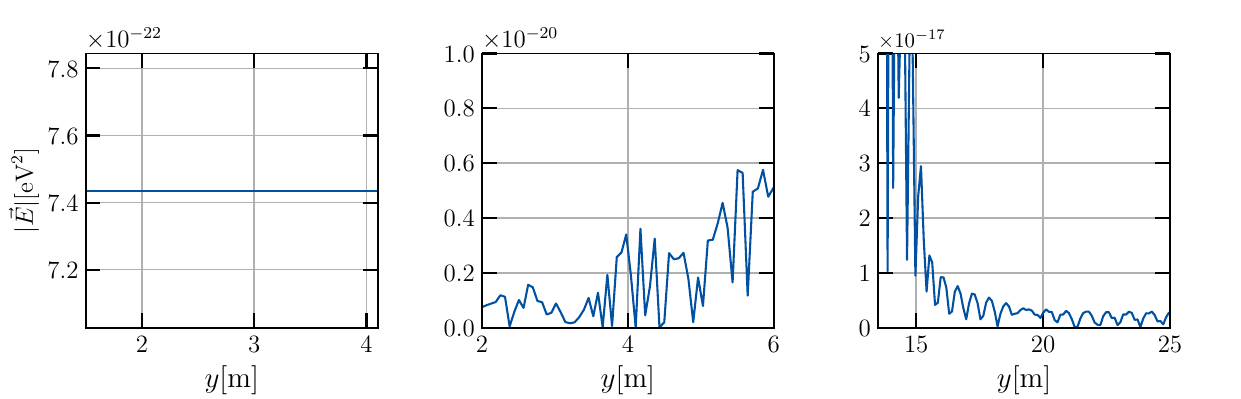"}	
        \caption{Reconstruction of refractive index. The top figure shows the reconstructed streamlines which describe the flow of the momentum $\vec{k}_\gamma$ over the simulation domain. The streamlines are colored by their refractive index. For the three highlighted streamlines we numerically compute the refractive index using Eq.~\ref{eq:wavevector}, and compare this to the dispersion relation of the associated mode, which is computed using the properties of the medium (see text for details). We identify trajectories 1, 2, and 3 as being associated with the magnetosonic-t, Alfv\'{e}n, and LO modes respectively. In the case of the Alfv\'{e}n mode, the large discrepancy arises from small errors in the reconstructed value of $\theta_{kB}$ -- this is due to the dispersion relation being situated near a pole, and therefore tiny variations in $\theta_{kB}$ translate into large errors in the reconstructed value of $n = |\vec{k}_\gamma|/\omega$. This point is illustrated directly in the plot by perturbing the reconstructed value of $\theta_{kB} \rightarrow \theta_{kB} \pm 3^\circ$, and plotting the perturbed refractive index using vertical `error' bars (these are of course not true error bars, but are merely being used to reflect the strong sensitivity of this reconstruction to numerical errors). }
	\label{fig:disp_rel}
\end{figure*}

The various electromagnetic modes that appear in magnetized plasmas can be differentiated via their respective dispersion relations. In order to identify which mode exists at a single point in the simulation domain we perform the following analysis. We first extract the momentum using Eq.~\ref{eq:wavevector}, which, together with the axion energy, gives us an estimate of the refractive index $n = |\vec{k}_\gamma| / \omega$. We then use the local value of $\vec{k}_\gamma$ to compute at each point $\theta_{kB}$ -- together with the local plasma frequency, this provides an alternative method for computing $n$. The computations are done along `streamlines', which are obtained by following the evolution of $\vec{k}_\gamma$ across the simulation domain. These streamlines are highlighted in Fig.~\ref{fig:disp_rel} (which uses parameters $v_a = 0.4$, $\theta_{B} = 0^\circ$, and $\theta_{\omega_p} = 20^\circ$), along with three examples of the reconstructed refractive index (shown along the trajectories labeled `1', `2', and `3'). The solid lines in the bottom panel are the numerical reconstruction (\ie the approach using only $|\vec{k}_\gamma|$ as extracted from Eq.~\ref{eq:wavevector}), while the dashed lines are those computed utilizing properties of the background fields.

The three trajectories identified in Fig.~\ref{fig:disp_rel} correspond to the magnetosonic-t, Alfv\'{e}n, and LO modes, respectively. Note that the reconstruction of the Alfv\'{e}n mode is numerically unstable -- this is because there exists a pole in the dispersion relation, and small variations in the reconstructed value of $\theta_{kB}$ can therefore lead to sizable differences in the inferred value of $\omega$. This can be illustrated by introducing an artificial error in the inferred value of $\theta_{kB}$ on the order of $\theta_{kB} \rightarrow \theta_{kB} \pm 3^\circ$ -- the impact of such a procedure on the reconstructed refractive index is shown using vertical error bars in Fig.~\ref{fig:disp_rel}. In the bottom panel of Fig.~\ref{fig:disp_rel} we also plot the amplitude of the local electric field associated to each mode -- here, one can see that the LO mode is produced with a much larger amplitude than the others, which is reflective of the fact that the axion-induced electric field in a dense plasma with $\omega < \omega_p$ should experience screening (see \eg~\cite{Caputo:2023cpv}). We note that the excitation of these modes appears to be physical, in that their amplitude scales perfectly with the axion-photon coupling, and their appearance and properties appear robust to \eg the resolution of the simulation.

For the sake of completeness, we include in the Appendix general streamline plots illustrating the behavior of the group velocity and the phase velocity over the simulation domain for the full suite of simulations.

\section{Conclusions}\label{sec:conclusions}
We have presented a numerical analysis of axion-photon mixing, demonstrating that resonant photon production can be accurately reconstructed in a variety of contexts. We have focused primarily on the case of resonant mixing in a strongly magnetized plasma, which is relevant for axion searches in neutron star magnetospheres. This has been an area of extensive research over the past few years, with multiple seemingly discrepant results appearing across the literature (despite appearing to rely on the same underlying assumptions). Our calculations show strong agreement across the entirety of parameter space with the most recent derivation performed in~\cite{mcdonald2023axionphoton}, thereby resolving outstanding questions about the efficiency of resonant mixing near neutron stars. This work thus represents an important step in solidifying the predictions and constraints derived on axion-photon mixing in these systems.

The framework developed here is rather general and can be applied to a wide variety of problems in axion electrodynamics, including scenarios in which the conventional assumptions required to obtain analytic predictions (such as \eg a slow variation in the properties of the background) break down, or to study the production of local electromagnetic modes in non-trivial backgrounds. Future work exploiting the ideas developed here will focus on axion-photon mixing in alternative environments, and the mixing of photons with other light fundamental particles such as dark photons.

\begin{acknowledgments}
The authors would like to thank Olga Mena for her contributions to the initial stages of this work. The authors would like to thank Jamie McDonald, Pete Millington, Bjorn Garbrecht, and  M.C. David Marsh for their comments on the draft. DN and CW are supported by the European Research Council (ERC) under the European Union's Horizon 2020 research and innovation programme (Grant agreement No. 864035 - Undark). SJW acknowledges support from a Royal Society University Research Fellowship (URF-R1-231065). This article/publication is based upon work from COST Action COSMIC WISPers CA21106, supported by COST (European Cooperation in Science and Technology).
This work has been partially supported by the Spanish MCIN/AEI/10.13039/501100011033 grants PID2020-113644GB-I00 and by the European ITN project HIDDeN (H2020-MSCA-ITN-2019/860881-HIDDeN) and SE project ASYMMETRY (HORIZON-MSCA-2021-SE-01/101086085-ASYMMETRY) as well as by the Generalitat Valenciana grants PROMETEO/2019/083 and CIPROM/2022/69. EUG acknowledges the financial support from the MCIU with funding from the European Union NextGenerationEU (PRTR-C17.I01) and Generalitat Valenciana (ASFAE/2022/020).
\end{acknowledgments}


\bibliography{main}

\begin{figure*}
	\centering
    \includegraphics[width=\linewidth] {"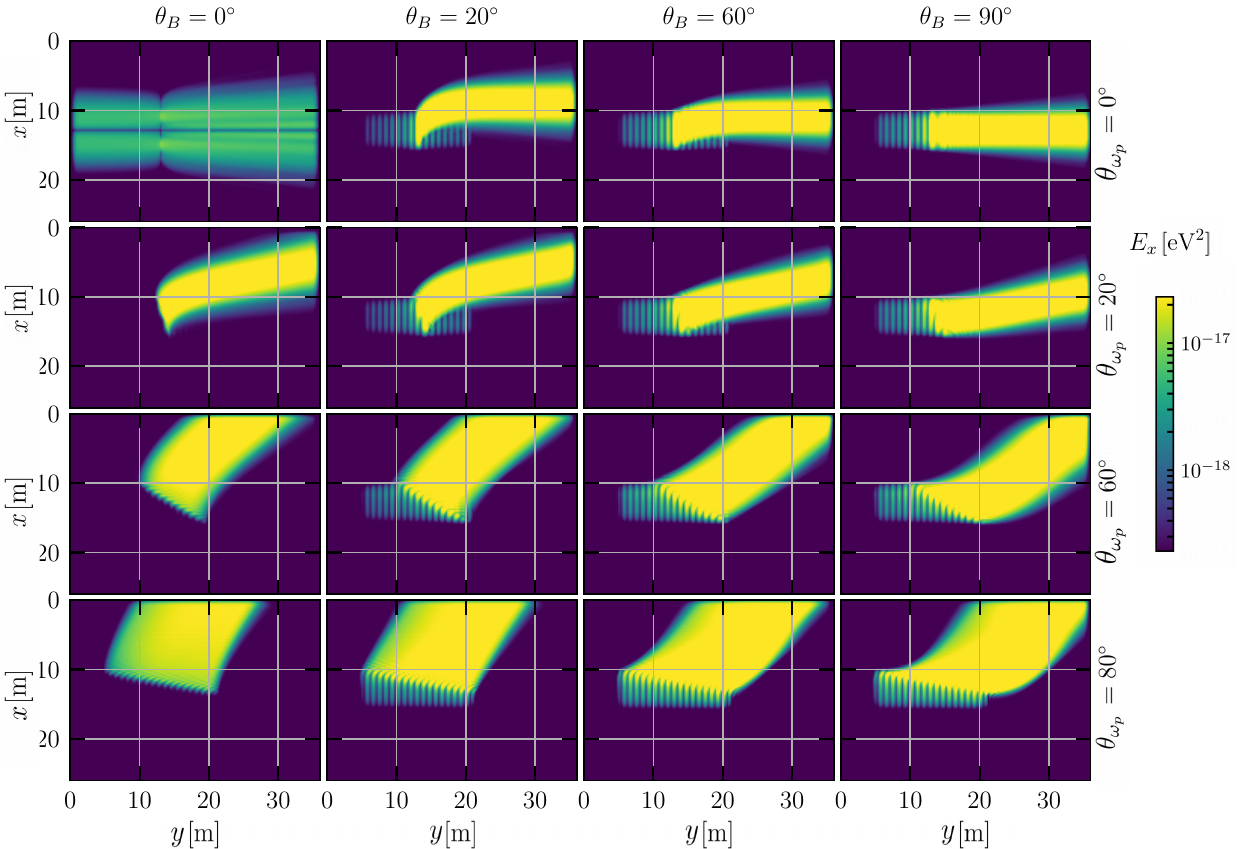"}	
	\includegraphics[width=\linewidth] {"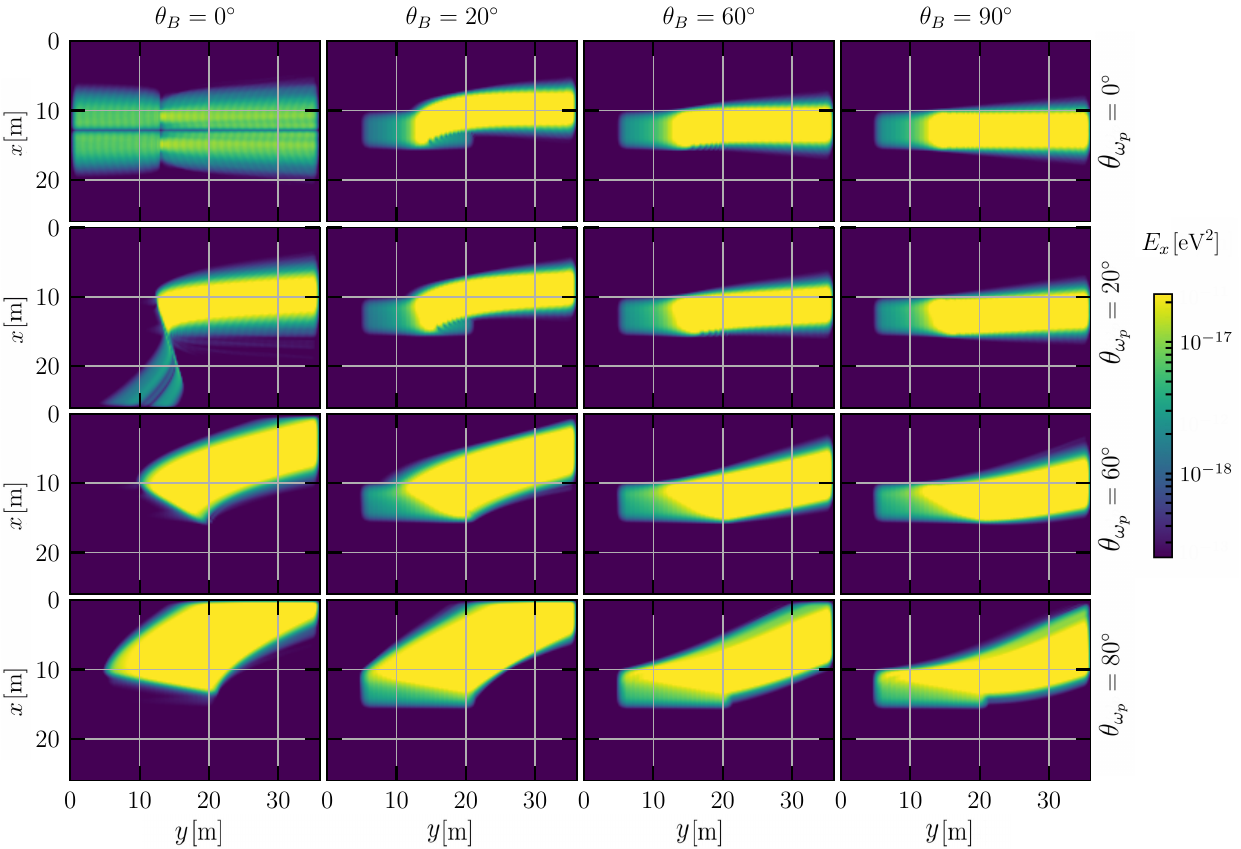"}	
	\caption{{\it Top:} Magnitude of the $E_x$ component for $v_a = 0.4$. {\it Bottom:} The same for $v_a = 0.8$. We plot for various choices of $\theta_{B}$ and $\theta_{\omega_p}$. }
	\label{fig:Ex_th_ph}
\end{figure*}

\begin{figure*}
	\centering
    \includegraphics[width=\linewidth] {"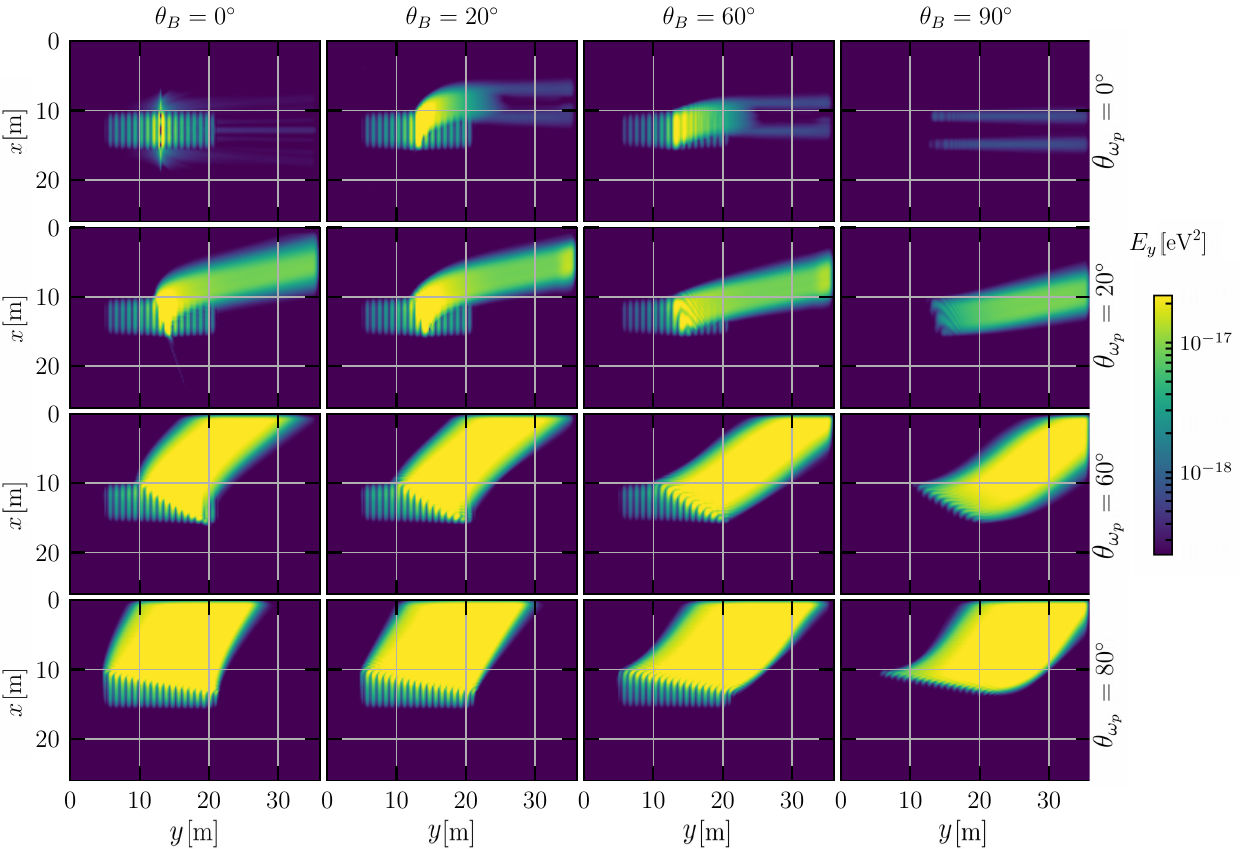"}	
	\includegraphics[width=\linewidth] {"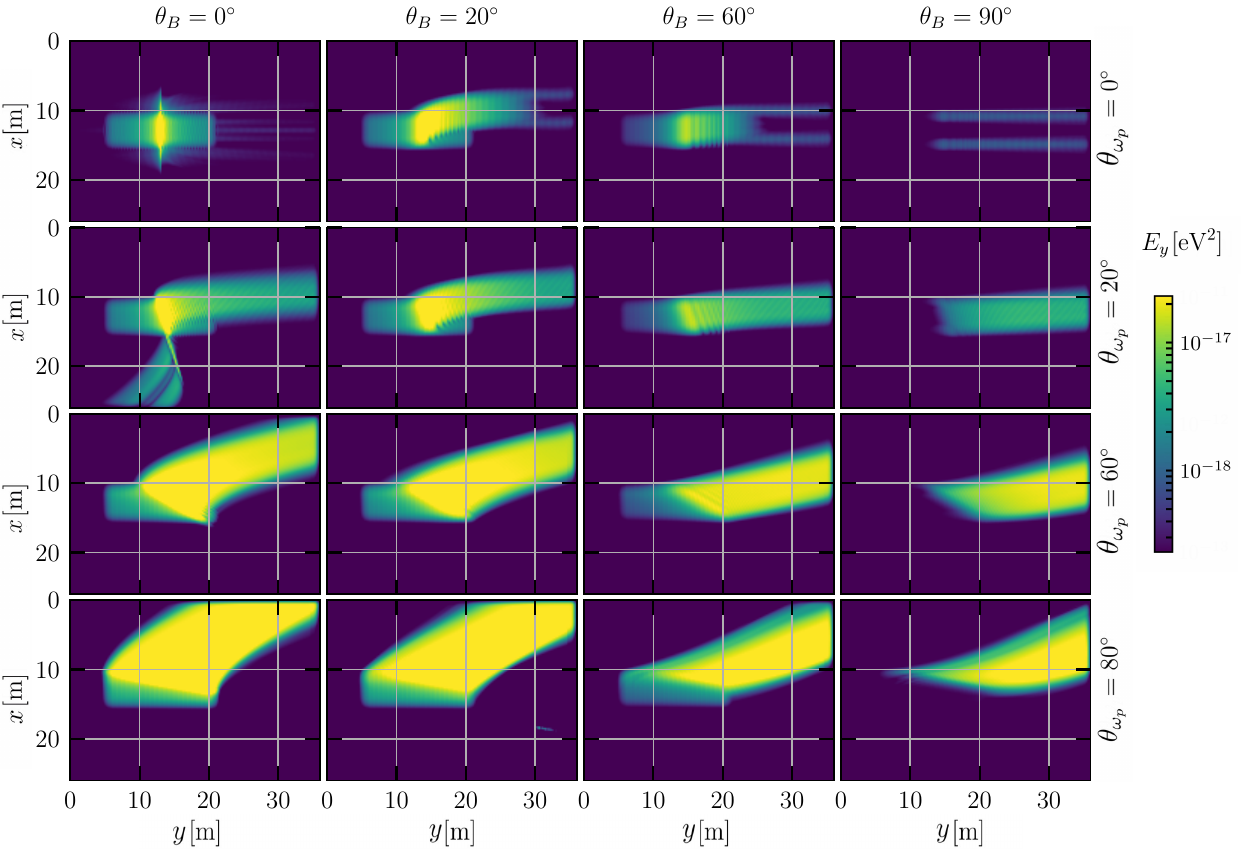"}	
	\caption{{\it Top:} Magnitude of the $E_y$ component for $v_a = 0.4$. {\it Bottom:} The same for $v_a = 0.8$. We plot for various choices of $\theta_{B}$ and $\theta_{\omega_p}$. }
	\label{fig:Ey_th_ph}
\end{figure*}

\begin{figure*}
	\centering
    \includegraphics[width=\linewidth] {"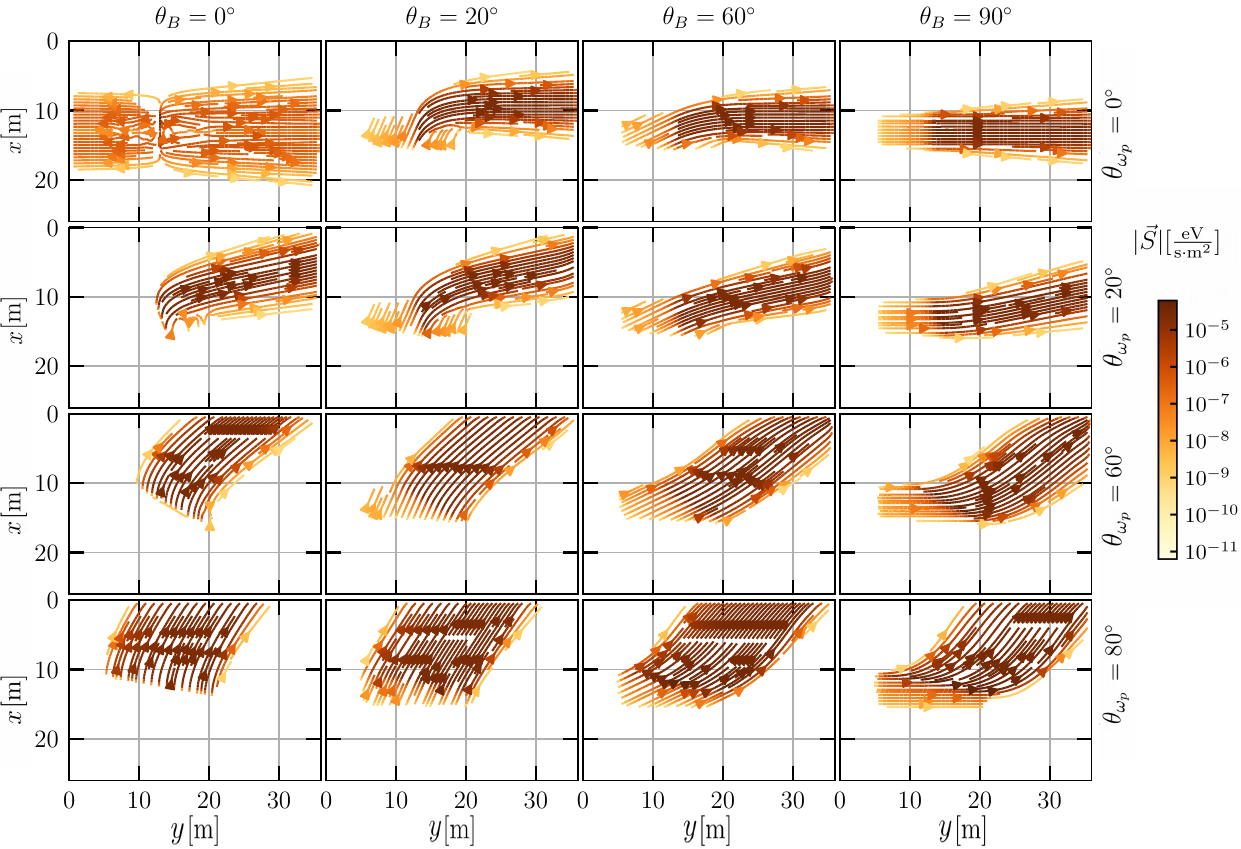"}	
    \includegraphics[width=\linewidth] {"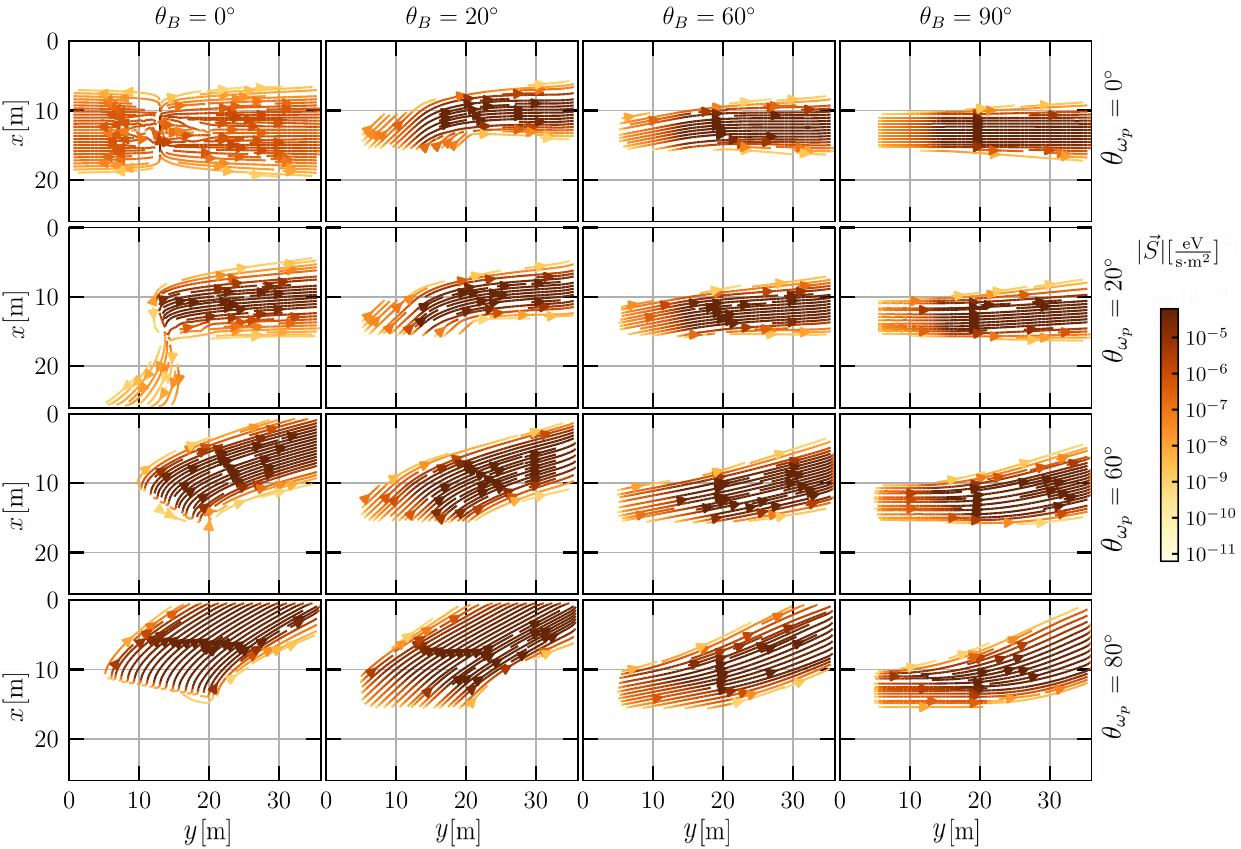"}	
	\caption{{\it Top:} Streamlines of the group velocity, with the magnitude being equal to the Poynting vector, for $v_a = 0.4$. {\it Bottom:} The same for $v_a = 0.8$. We plot for various choices of $\theta_{B}$ and $\theta_{\omega_p}$. }
	\label{fig:group_vel_th_ph}
\end{figure*}

\begin{figure*}
	\centering
    \includegraphics[width=\linewidth] {"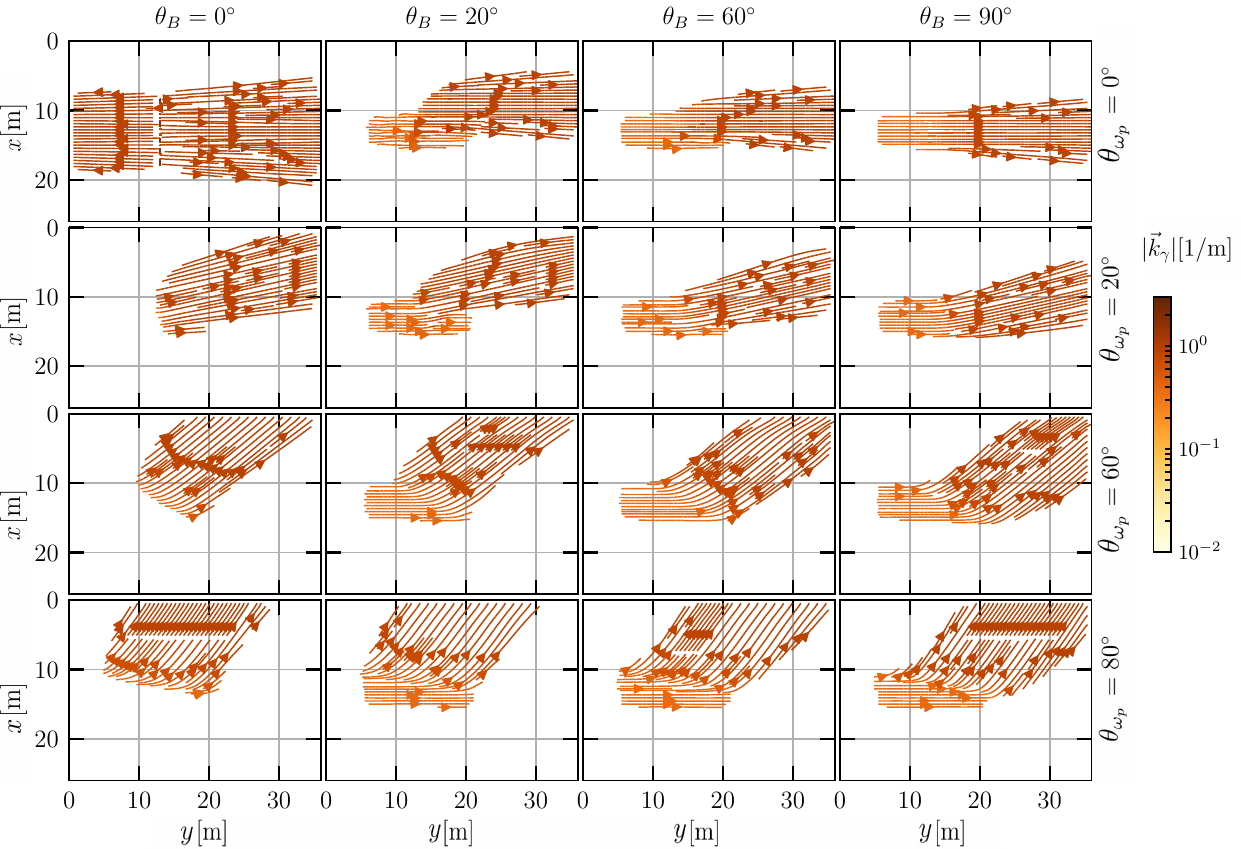"}	
	\includegraphics[width=\linewidth] {"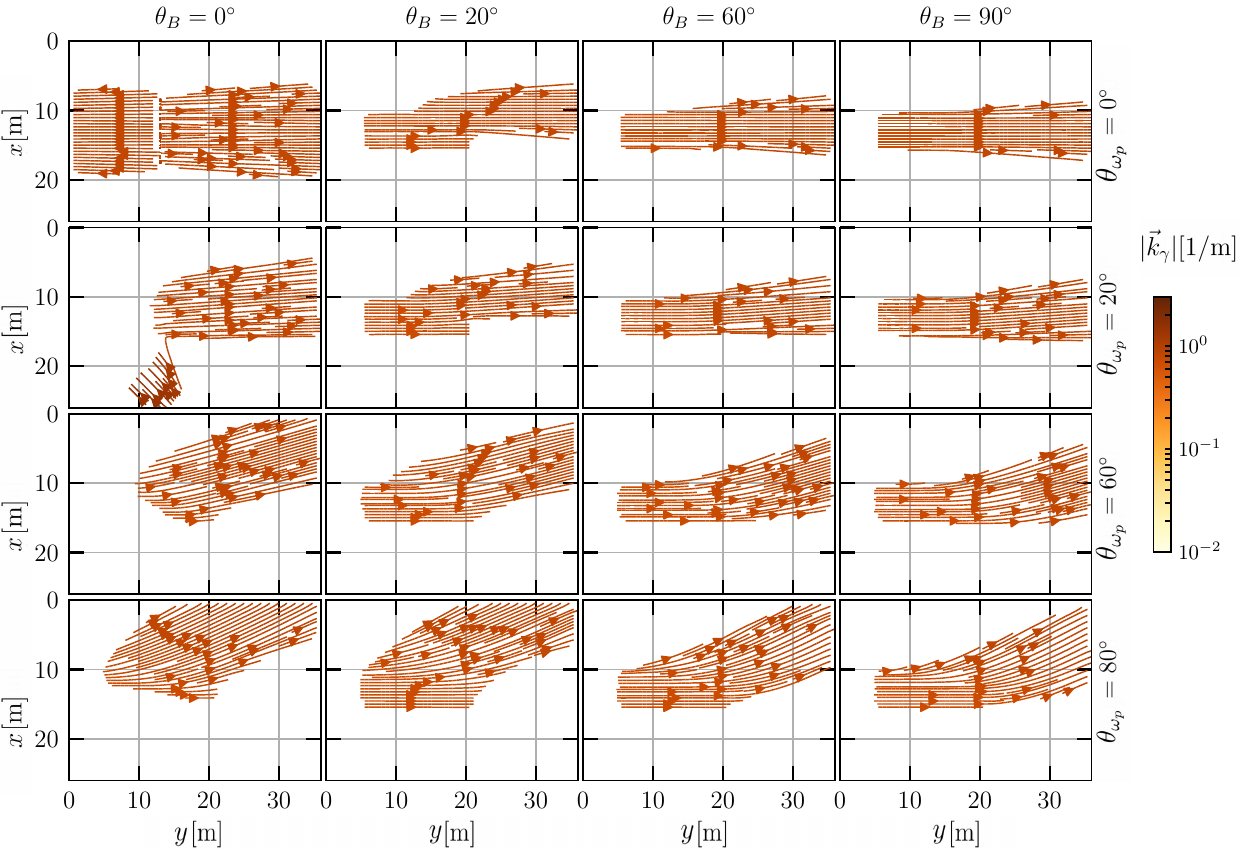"}	
	\caption{{\it Top:} Streamlines of the phase velocity, with the magnitude being equal to the momentum, for $v_a = 0.4$. {\it Bottom:} The same for $v_a = 0.8$. We plot for various choices of $\theta_{B}$ and $\theta_{\omega_p}$. }
	\label{fig:phase_vel_th_ph}
\end{figure*}

\section*{Appendix: Additional Numerical Results}
In this section we provide a closer look at various aspects of our numerical results, illustrating the behavior of our calculations across the entirety of the simulation domain, and showing the extent to which numerical convergence is achieved in our fiducial results.

We begin by showing the $x$- and $y$-components of the electric field (see Figs.~\ref{fig:Ex_th_ph} and~\ref{fig:Ey_th_ph}) across the entirety of the simulation domain for $v_a = 0.4$ (top) and $v_a = 0.8$ (bottom), four values of $\theta_{\omega_p}$, and four values of $\theta_B$. A variety of features can be directly seen in this plot, including the relative amplitude of each electric field component for each set of angles, photon refraction, and the axion-induced electric field. This figure illustrates the need for high resolution near the resonant point, as the amplitude of the electric field changes abruptly here. We also show in Fig.~\ref{fig:group_vel_th_ph} the Poynting vector, which more clearly illustrates the strong refraction observed in the previous plots. Finally, for the sake of completeness, we also show the streamlines of the reconstructed momentum across the simulation domain in Fig.~\ref{fig:phase_vel_th_ph}. These latter two figures illustrate the initial offset between the phase and group velocity at the point of conversion, as well as the evolution and convergence of these features as the wave propagates away from the plasma.  

\begin{figure*}
	\centering
    \includegraphics[width=1\linewidth] {"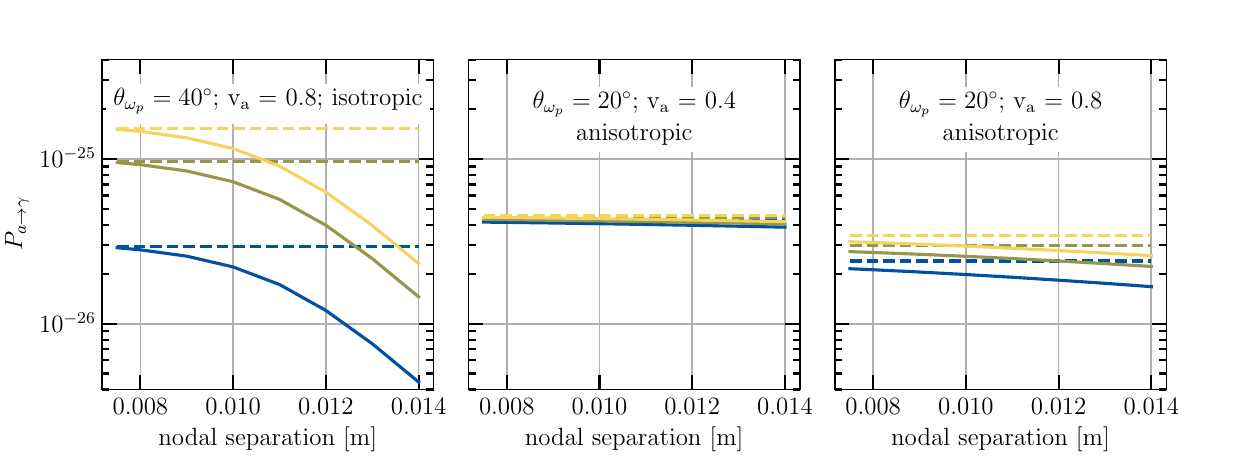"}
    \caption{Resolution sensitivity of the conversion probability at $\theta_B$ angles equal to $25^\circ$, $50^\circ$ and $75^\circ$ (blue, green and yellow lines). We compare the numerical (solid) results to the theoretical (dashed) values. The isotropic scenario (left) illustrates the nodal separation required to achieve the convergent behavior seen in Fig.~\ref{fig:pa_alpha}. The center and right panels illustrate the anisotropic scenario with $v_a = 0.4$ and $v_a=0.8$ respectively. Here we see that convergence occurs much quicker than in the isotropic case, and that the slight underestimation of the numerical results seen in Fig.~\ref{fig:pa_th_ph_v_0.4_0.8} is likely due to the slow convergence observed in the right panel. }
	\label{fig:resolution_convergence}
\end{figure*}

We finally turn our attention toward the issue of numerical convergence. We had argued in the main text that the slight discrepancies observed in the anisotropic conversion probability at $v_a = 0.8$ were a result of not being able to compute the mixing process at sufficiently high resolution. In order to illustrate this, we have computed the conversion probabilities for fixed values of $\theta_{\omega_p}$ and $v_a$, and varying the nodal separation from $\sim 7.5 \, \rm mm$ (our fiducial value for runs presented in the main text) to $14 \, \rm mm$; three examples are shown in Fig.~\ref{fig:resolution_convergence} for values of  $\theta_B$ set to $25$, $50$, and $75$ degrees. Here, one can see that the anisotropic conversion probability with low velocity is heavily converged, varying very little with nodal separation. The isotropic conversion probability shows a very strong convergence, but only achieves convergence as we approach our highest resolution. The anisotropic conversion probability at large velocity exhibits a much slower scaling, suggesting the current calculations slightly underestimate the true value of the conversion probability.

\end{document}